\documentclass[12pt]{article}

\title{ \vspace*{-0.3in}  \textbf{\Large Assortative Marriage and Geographic Sorting}\thanks{We thank Ying Jiang, Junrui Lin, Jingzhi Xu, Congying Yuan for their excellent research assistance. We also thank Jeremy Greenwood, Mark Rosenzweig, Jipeng Zhang, Junfu Zhang, as well as seminar participants at the GRIPS-UT macroeconomics workshop for their very helpful comments. This work was supported by the Natural Science Foundation of China [Grant No. 72133004].} }

\author{Jiaming Mao \thanks{Wang Yanan Institute of Studies in Economics (WISE), Xiamen University. Email: \href{mailto:jmao@xmu.edu.cn}{jmao@xmu.edu.cn}.} 
  \and{Jiayi Wen \thanks{School of Economics and Wang Yanan Institute of Studies in Economics (WISE), Xiamen University. E-mail: wjyecon@gmail.com.}}
}

\date{\small{ January, 2025} }

\usepackage{booktabs}
   \usepackage{natbib}
   \usepackage{amsmath}
\usepackage{bbm}   
   \usepackage{indentfirst}
   \usepackage{graphicx} % leave off '[demo]' in real document
   \usepackage{float}
   \usepackage{caption}
\usepackage[margin=1in]{geometry}
   \usepackage[justification=centering]{caption}
   \usepackage[normalem]{ulem}
   \usepackage[toc,page,title,titletoc,header]{appendix}
   \usepackage{multirow}
   \usepackage{epstopdf}

   \usepackage{booktabs}
   \usepackage{mathtools}
   \usepackage{rotating}
   \usepackage[colorlinks=true,linkcolor=blue,urlcolor=black,bookmarksopen=true]{hyperref}
   \usepackage{bookmark}
   \usepackage[titletoc]{appendix}   % package for adding appendix
   \usepackage{soul}    % package for strikethrough font
   \usepackage{setspace}
   \usepackage{numprint}

\usepackage{chngcntr} % For resetting counters
\usepackage{apptools} % For appendix customization
\AtAppendix{\counterwithin{table}{section}}
\AtAppendix{\counterwithin{figure}{section}}

\usepackage{titlesec}

\usepackage[table,xcdraw]{xcolor}   

\usepackage{array}
\usepackage{multirow}

\npdecimalsign{.}
\nprounddigits{3}

\hypersetup{
  colorlinks,
  citecolor=[RGB]{0,102,204},
  linkcolor=[RGB]{0,0,0},
  urlcolor=black}

\makeatletter

\newcommand{\Rmnum}[1]{\expandafter\@slowromancap\romannumeral #1@}
\makeatother

\usepackage[T1]{fontenc}
\usepackage[latin9]{inputenc}
\usepackage{geometry}
\usepackage{color}
\usepackage{amsmath}
\usepackage{amsthm}
\usepackage{amssymb}
\usepackage{setspace}

\begin{document}
\maketitle

\setstretch{1.5}

%%%%%%%%%%%%%
% The Paper %
%%%%%%%%%%%%%

\begin{abstract}
Between 1980 and 2000, the U.S. experienced a significant rise in geographic sorting and educational homogamy, with college graduates increasingly concentrating in high-skill cities and marrying similarly educated spouses. We develop and estimate a spatial equilibrium model with local labor, housing, and marriage markets, incorporating a marriage matching framework with transferable utility. Using the model, we estimate trends in assortative preferences, quantify the interplay between marital and geographic sorting, and assess their combined impact on household inequality. Welfare analyses show that after accounting for marriage, the college well-being gap grew substantially more than the college wage gap. \\

\textit{Keywords: assortative mating, spatial sorting, household inequality, college welfare gap, local marriage market, spatial equilibrium model}

%JEL code:  D10, J12, R23, 

\end{abstract}

%%%%%%%%%%%%%%%%
% Introduction %
%%%%%%%%%%%%%%%%
\newpage
\section{Introduction}
In recent decades, the U.S. has undergone significant changes in marriage and migration patterns, characterized by a marked rise in geographic sorting and educational homogamy. Increasingly, college graduates sort into high-skill cities \citep{moretti2012new} and marry similarly educated partners \citep{schwartz2005trends}. These shifts in where people choose to live and whom they choose to marry could have profound impact on household inequality and the welfare gap between college and non-college graduates.

In this paper, we argue that marital and geographic sorting are not independent. As college graduates increasingly reside in high-skill cities, the likelihood of marrying someone with a college degree increases, even without shifts in marital preferences. Thus what appears to be marital sorting could be driven by geographic sorting. Conversely, a preference for assortative marriage could draw unmarried college graduates to high-skill cities, where they are more likely to find suitable partners. In this way martial sorting could reinforce geographic sorting. 

To analyze the interplay between these phenomena and their implications for household inequality and the college welfare gap, we develop and estimate a spatial equilibrium model that incorporates local labor, housing, and marriage markets, integrating it with the marriage matching framework of \citet{choo2006marries}. Individuals in the model choose both where to live and work, and whether to marry and with whom, taking into consideration local wages, rents, and amenities. Together, these factors determine the spatial allocation of skills and marriage patterns that help explain the observed trends of assortative marriage and geographical sorting in the United States.

A key feature of our model lies in its assumption of local marriage markets. Much of the existing research on assortative mating and its economic consequences focuses on disentangling assortative preferences from shifting national education distributions, implicitly assuming a national marriage market \citep{greenwood2014marry, chiappori2017partner, eika2019educational, gihleb2020educational}. However, marriage is more likely to be a local affair. People tend to form relationships and marry spouses who reside in close proximity to them due to a range of factors such as work relationships, shared social networks, and geographic convenience. Empirically, we show substantial variation in marriage patterns across cities, underscoring the importance of this local dimension. Ignoring the local nature of marriage markets can lead to a biased understanding of how people marry assortatively. For instance, consider a nation with two cities, A and B: Suppose most college graduates live in A and marry randomly within their city, while most non-college graduates live in B and do the same. Under a national marriage market assumption, we would wrongly conclude that people have a strong assortative preference---when no such preference exists. Moreover, as college graduates migrate from B to A over time, this geographic sorting naturally increases educational homogamy at the national level, which could be misinterpreted as a consequence of rising assortative preference. Therefore, accounting for local marriage markets is essential for accurately understanding observed marriage patterns and its economic consequences.

We estimate the model using U.S. Census data from 1980 to 2000, a period marked by significant changes in labor and marriage markets \citep{diamond2022spatial}. While the transferable utility matching framework of \citet{choo2006marries} is just-identified, our extension to a local market setting enables identification of marital preference parameters through spatial variations in marital outcomes, wages and skill compositions, allowing us to uncover marital surpluses at the local level and derive novel findings. Following \citet{diamond2016determinants}, parameters related to labor and housing markets are estimated using local labor demand shocks and land availability as key instruments, with exposure to national industrial shocks serving as shift-share instruments to capture exogenous variation.

Based on the estimated model, our analysis delivers five set of results. First, we analyze trends in assortative preference, using a measure of assortative preference based on the supermodularity of marital surplus recovered by our structural model. We document the existence of a strong preference for assortative marriage, yet this preference slightly declined between 1980 and 2000, suggesting that the rise in educational homogamy was not driven by an increase in assortative preference. Then, we demonstrate that assuming a national marriage market---in a period with notable geographic sorting---would overestimate the changes in assortative preferences by almost 27 percent, highlighting the importance of considering local marriage markets. Finally, we decompose the overall preference into its pecuniary and nonpecuniary components, finding that both contributed to the decline in assortative preferences over time. 

Second, we explore the determinants of changes in marriage matching patterns over time, focusing on the gap between college and non-college graduates in their probability of marrying a high-skill spouse (hereinafter PMH). Between 1980 and 2000, this college marital gap widened by 6 percent, reflecting increased educational homogamy. Our analysis shows that this increase was primarily driven by the secular rise in educational attainment, particularly among women. However, assortative preference plays a critical role despite its own decline during this period. The gap widened because rising educational attainment interacted with positive assortative preferences---As the college-educated population grew, the preference for marrying similarly educated spouses disproportionately increased the likelihood of college graduates marrying one another, thus amplifying the college marital gap.

In addition to the secular rise in educational attainment, the process of geographic sorting further contributed to the widening college marital gap. Increasingly, college and non-college graduates face different marital pools due to their diverging location choices. This led to an increase in the number of assortatively matched households nationwide. Overall, we show that the share of couples that are both college-educated in the U.S. rose as a result of geographic sorting. Even without assortative preferences, college graduates had a 1.6 percentage point higher PMH than non-college graduates in 1980, driven by differences in their location choices---a gap that widened to 2.4 percentage points by 2000.

Third, we explore the role of marital prospects in influencing the location choice of college and non-college graduates. We find that while marriage \textit{per se} leads to a more even spatial allocation of skills, a finding consistent with \cite{Fan2022The}, a stronger preference for assortative marriage would actually increase the spatial inequality in educational levels. This is because the prospect of finding high-skill partners would increase the relative value of a high-skill city to unmarried college graduates. Assortative preferences would therefore amplify the process of geographic sorting. Thess findings also have two implications over time. On the one hand, it suggests that the weakening role of marriage have contributed to the observed rise in geographic sorting. On the other hand, the declining assortative preference has conversely ameliorated this process. 

Fourth, we investigate the implications of marital and geographic sorting on household inequality. We find that marital assortativeness amplified national inequality by 4.9 percent in 1980, and the impact increased to 7.4 percent in 2000. Despite its decline, marital assortativeness therefore has contributed to roughly 4.2 percent of the overall rise in national inequality. We show that, during this period, nationwide increases in educational attainment and skill premiums interacted with assortative preferences to form more high-skill, high-income households. Additionally, geographic sorting led to more assortatively matched households concentrating in high-skill cities. Over time, these cities also experienced faster productivity and wage growth, further amplifying national inequality. We find that the combined effects of geographic sorting and uneven wage growth accounted for over 4 percent of the observed growth in national inequality. 

Across cities, local education composition and wage structure also interacted with assortative preferences to make some cities more unequal than others. In high-skill cities with high college premiums, marital assortativeness contributed the most to raising local household inequality. 

Fifth, we analyze the welfare implications of individual location and marital choices. Based on the structural model, we recover and then decompose local marital surpluses into a pecuniary component, a nonpecuniary component, and marital transfers. Our analysis reveals important welfare effects through intrahousehold marital transfers under marriage market equilibrium. For instance, although males can receive higher pecuniary benefits in a city with higher female wages, they also need to pay more marital transfers.  Overall, we reveal that high-skill individuals
benefit less from marriage by living in high-skill cities, explaining our findings that marriage leads to less dispersive skill distributions across cities.

Overtime, as college and non-college graduates increasingly reside in different cities with different wages and rents, we find a widening college well-being gap quantitatively similar to \cite{moretti2013real} and \cite{ diamond2016determinants}. Moreover, we find that this gap caused by real wage changes was further amplified through intrahousehold marital transfers.
Shifting education distributions also have important welfare effects through this channel: As the share of high-skill females have experienced the most significant increase, their welfare has increased by the less extent due to less marital transfers they receive. In contrast, the college well-being gap among males become further larger because this change generates more transfers for high-skill males under marital assortativeness .

\paragraph{Literature} Our paper is broadly related to two strands of literature: the labor economics research on marriage and its implications for inequality, and the urban economics research on location choices and their welfare consequences. We make several contributions. First, we contribute to the literature on trends in assortative mating, which has traditionally focused on disentangling assortative preferences from shifting educational distributions \citep{greenwood2014marry, siow2015testing, chiappori2017partner, eika2019educational, gihleb2020educational, chiappori2020changes}. Unlike most existing studies, which predominantly assume a national marriage market, we hypothesize and empirically demonstrate the upward bias introduced by this assumption under geographic sorting. By employing a structural model, we measure assortative preferences through the estimated supermodularity core of marital surpluses and disentangle the role of pecuniary and nonpecuniary factors.

Second, we contribute to the literature on the inequality implications of assortative marriage \citep{fernandez2005love, greenwood2014marry, greenwood2016technology, eika2019educational}. Our findings demonstrate how rising educational attainment and skill premiums interacted with assortative preferences to exacerbate household inequality at both the national and local levels, and how geographic sorting further amplified this effect. Moreover, by examining the role of marital assortativeness in shaping local inequality, we contribute to broader discussions on the determinants and consequences of spatial inequality \citep{glaeser2009inequality, Glaeser2016UnhappyCities}.

Third, we contribute to the understanding of spatial heterogeneity in marital outcomes. While we are not the first to examine local marriage markets \citep{angrist2002sex, abramitzky2011marrying, grosjean2019s}, much of the existing literature primarily focuses on the role of local sex ratios. Our framework goes beyond this by incorporating the supply and demand of marital candidates based on education and by endogenizing local marital pools through individuals' location choices within a spatial equilibrium framework. This approach allows us to uncover novel findings on the spatial patterns in marital preferences, marital choices, and their inequality consequences.

In addition to contributions to family and labor economics, our paper enriches the urban economics literature on the causes and consequences of geographic sorting. A series of studies, including \cite{blanchard1992regional}, \cite{notowidigdo2020incidence}, and \cite{hornbeck2022estimating}, emphasize how local productivity shocks drive individuals' location choices\footnote{See \cite{Moretti2011LocalMarkets} for a comprehensive review.}, while \cite{diamond2016determinants} highlights the importance of local amenities. Our study adds to this literature by introducing and quantifying the role of marital prospects in shaping individual location choices and influencing the spatial sorting of individuals.

On the well-being impacts of geographic sorting,  \cite{moretti2013real} shows that the college welfare gap has grown less than the observed college wage gap after accounting for local rents, while \cite{card2023location} suggests that higher rents effectively offset the wage advantages of college graduates. In contrast, \cite{diamond2016determinants} finds that, when local amenities are considered, the welfare gap between college and noncollege workers has widened significantly more than their wage gap. This paper extends this line of analysis by examining the role of marriage, showing that marital surplus and geographic sorting together contribute to a larger college welfare gap than can be explained by wages, rents, and amenities alone.

Finally, this paper situates itself within the emerging literature studying the interaction between local labor and marriage markets with quantitative spatial models \citep{Fan2022The, alonzo2023segregation, alonzo2025marrying}. The focus of these studies is primarily on how marriage affects spatial distributions of skills. Our paper is motivated by how geographic and marital sorting along the education dimension affect each other. While the impact of marriage on location choices is part of our findings, another important perspective of the current paper is how missing geographic sorting and local marriage markets can bias our understanding on assortative mating and its implication for household inequality. We also provide welfare analysis building on the literature focusing on college well-being gap.\footnote{Technically, this paper and \cite{Fan2022The} share the most similar framework, both innovatively and independently integrating a marriage matching framework to a spatial model. However, the two papers have significantly different focuses. First, \cite{Fan2022The} use their model to study how marriage affects spatial allocation of economic activities, whereas we focus on the interaction between spatial and marital sorting. Second, we further discuss implications for inequality, adding to \cite{greenwood2014marry, greenwood2016technology, eika2019educational} and for college well-being gap, to \cite{Moretti2013RealInequality, diamond2016determinants}. Third, \cite{Fan2022The} calibrate their model, with a relatively richer setting, using Census 2000, while we estimate the model by both the Bartik shocks between 1980 and 2000, and the spatial variation across local markets.} 

The rest of this paper is organized as follows. In Section 2, we describe stylized facts regarding marriage and geographic sorting. Section 3 presents our model. Section 4 discuss identification and estimation methods. Section 5 presents estimation results and describes model fits. In Section 6, we use our model to analyze how assortative marriage and geographic sorting influence each other. Section 7 and 8 analyze the inequality and welfare consequences. Section 9 concludes this paper.

\section{Stylized Facts} \label{sec:desc}

Both education attainments and marriage outcomes have experienced remarkable changes during recent decades in the U.S. In 1980, the probability of marrying a high-skill spouse (PMH) was 17.3 percent for males and 28.0 percent for females. By 2000, the PMH for males had nearly doubled to 33.3 percent, while for females, it increased to 36.6 percent. This notable rise in the male PMH coincided with a 15.5-percentage-point increase in the share of college-educated females.

The shifting educational distributions, however, account for only part of the observed trends. Marital preferences also play a pivotal role in shaping marital outcomes. As the share of high-skill females increased, the PMH for high-skill males rose by 15.1 percentage points---nearly twice the 8.8-percentage-point increase observed for low-skill males. This disparity underscores the significant influence of assortative marital preference, as these increases would have been identical for both groups under random matching.

%The shifting educational distribution, however, tells only part of the story. Marital preferences also play a crucial role in shaping marital outcomes. As the share of high-skill females increased, the PMH for high-skill males rose by 15.1 percentage points---nearly twice as much as the 8.8 percentage point increase for low-skill males.  This fact indicates the significant influence of assortative marital preferences, since these increases would have been identical for both groups if matching were random.

Furthermore, there is substantial geographic heterogeneity in marital matching outcomes across cities. Table~\ref{tab:cityPMH} highlights the top and bottom five cities in PMH by education level, revealing pronounced  educational and spatial differences. For instance, in 1980, a college graduate in Ann Arbor, MI, had a 73 percent probability of marrying an equally educated spouse, while a non-college-educated individual in Danville, VA, faced just a 5 percent probability. Assortative preference on education strongly shape these patterns. Even in Washington, DC---ranked highest for PMH among non-college-educated individuals in 1980---only 23 percent of them married a college-educated spouse.

However, location also plays a critical role, as evidenced by the notable PMH gap  between Washington, DC and Danville, VA for non-college-educated individuals (0.23 v.s. 0.05). Similarly, the PMH for college graduates declines from 73 percent in Ann Arbor, MI, to just 41 percent in Johnstown, PA. By 2000, the PMH for non-college-educated individuals in the top-ranked city, Stamford, CT, stands at 31 percent, approaching the 42 percent PMH of college-educated individuals in the lowest-ranked city, Jacksonville, NC.

%Furthermore, there is large geographic heterogeneity in marital matching outcomes across cities. Table~\ref{tab:cityPMH} lists the top and bottom five cities in PMH by education level, revealing stark educational and spatial differences. For instance, in 1980, a college-educated individual in Ann Arbor, MI, had a 73 percent probability of marrying an equally educated spouse, while a non-college-educated individual in Danville, VA, faced only a 5 percent probability. Assortative preferences on education significantly shape these outcomes. Even in Washington, DC, which ranked highest in PMH for non-college-educated individuals in 1980, only 23 percent of them married a college-educated spouse. However, location also plays a significant role as evident by the PMH gap  between Washington, DC and Danville, VA for non-college-educated individuals. The PMH for college graduates also declined from 73 percent in Ann Arbor, MI, to only 43 percent in Lima, OH. Furthermore, by 2000, the PMH for non-college-educated individuals in the top city (Stamford, CT) stands at 31 percent, approaching the 42 percent probability of college-educated individuals in the lowest-ranked city (Jacksonville, NC).

% Table
\begin{table}[t]
{\scriptsize
\caption{\small{Probability of Marrying a High-Skill Spouse in Different Cities}}
\label{tab:cityPMH}
\resizebox{\textwidth}{!}{
\begin{minipage}[c]{0.99\textwidth}
\tabcolsep=0.01cm
\begin{tabular}{p{0.17\textwidth}p{0.08\textwidth}p{0.17\textwidth}p{0.08\textwidth}p{0.19\textwidth}p{0.08\textwidth}p{0.17\textwidth}p{0.05\textwidth}}
\toprule
\toprule
\multicolumn{4}{c}{\textbf{1980}}  & \multicolumn{4}{c}{\textbf{2000}}  \\
      \cmidrule(lr){1-4}    \cmidrule(lr){5-8}  
\multicolumn{2}{c}{\textbf{College}} & \multicolumn{2}{c}{\textbf{Noncollege}} & \multicolumn{2}{c}{\textbf{College}} & \multicolumn{2}{c}{\textbf{Noncollege}} \tabularnewline
\midrule
\multicolumn{2}{l}{\textit{Top Cities} }  & & & & & & \vspace{0.2cm}\tabularnewline
Ann Arbor, MI & 0.73 & Washington, DC & 0.23 & Stamford, CT & 0.84 & Stamford, CT & 0.31 \tabularnewline
Columbia, MO & 0.69 & Columbia, MO & 0.22 & Trenton, NJ & 0.79 & Gainesville, FL & 0.27 \tabularnewline
Madison, WI & 0.68 & Stamford, CT & 0.20 & San Jose, CA & 0.78 & Fort Collins, CO & 0.25 \tabularnewline
Lafayette, IN & 0.68 & Provo, UT & 0.19 & Columbia, MO & 0.77 & Ann Arbor, MI & 0.25 \tabularnewline
State College, PA & 0.68 & San Jose, CA & 0.19 & Washington, DC & 0.77 & Washington, DC & 0.25 \vspace{0.15cm} \tabularnewline
\midrule
\textit{Bottom Cities} & & & & & & & \vspace{0.2cm}\tabularnewline
Lima, OH & 0.43 & Lima, OH & 0.06 & Johnstown, PA & 0.49 & Yakima, WA & 0.09 \tabularnewline
Waterbury, CT & 0.43 & Hagerstown, MD & 0.06 & Daytona Beach, FL & 0.48 & Brownsville, TX & 0.09 \tabularnewline
Manchester, NH & 0.43 & Brownsville, TX & 0.06 & Altoona, PA & 0.47 & McAllen, TX & 0.08 \tabularnewline
Yakima, WA & 0.42 & Waterbury, CT & 0.06 & Vineland, NJ & 0.47 & Visalia, CA & 0.08 \tabularnewline
Johnstown, PA & 0.41 & Danville, VA & 0.05 & Jacksonville, NC & 0.42 & Danville, VA & 0.07 \tabularnewline
\bottomrule
\end{tabular} 

\vspace{0.5em}
\textit{Notes:} This table presents the probability of marrying a high-skill spouse (PMH) in the top and bottom five cities in 1980 and 2000. High-skill is defined as college-educated.
\end{minipage}
}
}
\end{table}
% end 

% Figure
\begin{figure}[t]
\centering
\begin{minipage}[c]{0.99\textwidth}
\includegraphics[height=6.3cm]%{DF1.eps}
{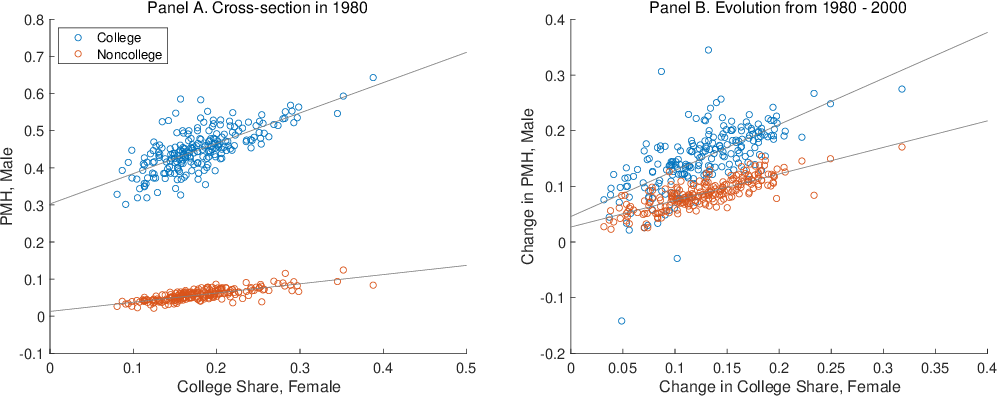}
\caption{Correlation between Marriage Outcomes  and Education Distributions } 

\vspace{1em}
{\scriptsize \textit{Notes:} This figure presents the scatter plots of the probability of marrying high-skill spouse (PMH) of males versus the share of college-educated females across cities. Panel A presents the cross-sectional variation in 1980. Panel B presents the over time changes from 1980 to 2000.}{\scriptsize\par}
\label{figure:df2}
\end{minipage}
\end{figure}
% end

% Figure
\begin{figure}[t]
\centering
\begin{minipage}[c]{0.95\textwidth}
\includegraphics[height=6cm]{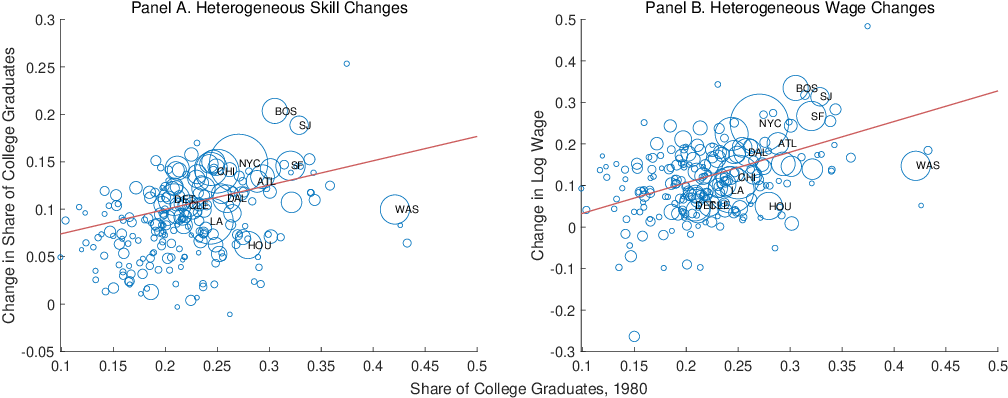}
\caption{Evolution of Skills and Wages across Cities by 1980 Education Level}

\vspace{1.0em}
{\scriptsize  \textit{Notes:} This figure presents the changes from 1980 to 2000 of the share of college graduates and the log wage across cities with different education level in 1980.}{\scriptsize\par}
\label{figure:df1}
\end{minipage}
\end{figure}
% end

Panel A of Figure~\ref{figure:df2} illustrates the cross-sectional variation in PMH in 1980, emphasizing the roles of local education distributions and marital preferences in shaping outcomes. Across cities, male PMH is strongly correlated with the local share of college-educated females. Moreover, even when faced with identical shares of college-educated females, high-skill males are significantly more likely to marry high-skill females compared to low-skill males---a pattern indicative of strong assortative preferences. Notably, this college marital gap widens substantially in cities with larger shares of high-skill females, suggesting an interaction between marital preferences and local education compositions in determining marital outcomes.

Over time, the factors influencing marriage have evolved differently across cities. Figure~\ref{figure:df1} demonstrates that college graduates increasingly concentrated in cities with already high proportions of college graduates in 1980. These cities also experienced greater wage growth, a trend termed the ``Great Divergence'' by \cite{moretti2012new}. Panel B of Figure~\ref{figure:df2} depicts the relationship between changes in male PMH and changes in the share of high-skill females, mirroring the positive cross-sectional correlation observed in 1980. Furthermore, cities with larger increases in the share of college-educated females exhibit a greater rise in PMH for high-skill males relative to low-skill males. This divergence underscores the dynamic interaction between marital preferences and educational compositions in shaping local marriage markets, where high-skill males are disproportionately more likely to marry high-skill females as the pool of college-educated females expands.

In conclusion, the evidence highlights substantial variation in marital outcomes across cities and over time, where both education distributions and assortative preferences seem to be key determinants. In the following section, we present our model, which formally captures the spatial heterogeneity and the interactions between these  factors.

\section{Model}

To capture the spatial distributions of wages, rents, and marriage matching outcomes, we build a matching model of marriage under transferable utility, building on \cite{choo2006marries}, and integrate it into a \cite{diamond2016determinants}-type spatial equilibrium model across cities. In the model, individuals make decisions at two stages, first choosing where to live, and then whether to marry and with whom. Labor, housing, and marriage markets are local.

When choosing a location, individuals consider not only local wages, rents, and amenities but also marital prospects. These prospects are shaped by expectations regarding the likelihood of marrying individuals of various types or remaining single, as well as the associated benefits.  Both the likelihood and benefits vary across cities due to differences in wage levels and education compositions.

Let individual $i$ be characterized by gender $g_{i}\in\{M,F\}$, where $M$ and $F$ denote male and female, and by education $e_{i}\in\{H,L\}$, where $H$ and $L$ represent high-skill and low-skill. The local markets in each city consist of the housing market, labor markets for two skill levels, and marriage markets encompassing four types of couples, differentiated by their educational combinations.

\subsection{Labor Market}

\subsubsection*{Labor Demand}

The production of tradable goods in city 
$m$ during period  $t$ follows a constant elasticity of substitution (CES) form:
\begin{equation}
P_{t}Y_{mt}=P_{t}A_{mt}\left[\alpha_{mt}\left(\mathcal{L}_{mt}^{H}\right)^{\rho}+\left(1-\alpha_{mt}\right)\left(\mathcal{L}_{mt}^{L}\right)^{\rho}\right]^{\frac{1}{\rho}},\label{eq:prod00}
\end{equation}
where $A_{mt}$ denotes Hicks-neutral technology. $\mathcal{L}_{mt}^{L}$ and $\mathcal{L}_{mt}^{H}$ are the aggregate low-skill and high-skill \emph{effective labor} employed in city $m$, receptively, and $\sigma=1/\left(1-\rho\right)$ represents the elasticity of substitution between high-skill and low-skill labor.  The exogenous productivity $A_{mt}$ and the high-skill labor share $\alpha_{mt}$ vary across cities and over time.  The price of final goods $P_{t}$ is normalized to $1$. The labor market is perfectly competitive, so each unit of skill-specific effective labor thus earns its marginal return: 
\begin{equation}
w_{mt}^{e}=\left.\partial Y_{mt}\right/\partial\mathcal{L}_{mt}^{e}, \text{ for } e \in \{ H, L \} . \label{eq:prod01}
\end{equation}

\subsubsection*{Effective Labor}

Our framework focuses on individuals' location and marriage decisions while abstracting from their labor supply choices. Nevertheless, individuals supply varying levels of effective labor, which depend on their gender and marital status. Specifically, each individual provides either $\ell_{it}^{H}$ units of high-skill effective labor or $\ell_{it}^{L}$ units of low-skill effective labor, determined by their education level:
\begin{align}
\ell_{it}^{H} & =\exp\left[\phi_{t}^{H}\cdot\mathbbm{1}(g_{i}=F)+\delta_{t}^{H}\cdot\mathbbm{1}(g_{i}=F)\cdot\mathbbm{1}(i\text{ is married})\right]\label{eq:labor000}\\
\ell_{it}^{L} & =\exp\left[\phi_{t}^{L}\cdot\mathbbm{1}(g_{i}=F)+\delta_{t}^{L}\cdot\mathbbm{1}(g_{i}=F)\cdot\mathbbm{1}(i\text{ is married})\right],\nonumber 
\end{align}
where $\phi_{t}^{e}$ reflects the gender gap in labor income, while married females supply less effective labor, subject to a  discount $\delta_{t}^{e}$, which depends on their education level $e\in\left\{ H,L\right\} $. $\delta_{t}^{e}$ is also  allowed to vary over time to reflect changes in married female labor force participation.
$\left(\delta_{t}^{e},\phi_{t}^{e}\right)$ are assumed to be exogenous, with the effective labor of males normalized to $1$ for both high-skill and low-skill workers.\footnote{Assuming $\left(\delta_{t}^{e},\phi_{t}^{e}\right)$ to be exogenous provides a parsimonious way to account for labor supply decisions and factors such as workplace discrimination that contribute to the observed gender wage gap. Modeling the gender wage gap as differences in effective labor units is common in labor economics. See, e.g., \cite{lee2005estimable}; \cite{lee2006intersectoral}.} 

The aggregate skill-specific labor supply in city $m$ is the sum of skill-specific effective labor provided by all workers of that type, where 
\begin{equation}
\mathcal{L}_{mt}^{e}=\sum_{i}\ell_{it}^{e}\cdot\mathbbm{1}\left(i\text{ lives in city }m\right)\label{eq:labor02}
\end{equation}

Individual wage income is determined by the skill rental price $w_{mt}$ and the effective labor $\ell_{it}^{e_{i}}$:
\begin{equation}
\ln W_{imt}=\ln w_{mt}^{e_{i}}+\ln\ell_{it}^{e_{i}}\label{eq:labor03}
\end{equation}

\subsection{Housing Market}

Our model of housing demand and housing supply follows that of \cite{diamond2016determinants}. Each household is assumed to spend $\zeta\in\left(0,1\right)$ of total income on housing and local goods.\footnote{We treat local goods and housing as equivalent since local goods prices are determined by local housing prices. Therefore, we will use \textquotedblleft local goods\textquotedblright{} and \textquotedblleft housing\textquotedblright{} interchangeably throughout the paper.} Let $\mathcal{W}_{mt}$ be the aggregate wage income in city $m$. Then the aggregate housing demand is captured by 
\begin{equation}
\mathcal{H}_{mt}=\frac{\zeta}{R_{mt}}\mathcal{W}_{mt},\label{eq:housing00}
\end{equation}
where $R_{mt}$ is the rent in city $m$. We follow the standard assumption in urban literature to assume absentee landlords
so that residents derive their income from labor earnings only. 

On the supply side, developers are price-takers in competitive housing markets and sell each unit of homogeneous houses at the marginal cost of production:
\begin{equation}
R_{mt}=\kappa_{t}\cdot \text{MC}\left(\text{CC}_{mt},\text{LC}_{mt}\right),\label{eq:housing02}
\end{equation}
where $\kappa_{t}$ is the annual interest rate. The marginal cost of housing $\text{MC}$ is a function of construction costs $\text{CC}_{mt}$ and local land costs $\text{LC}_{mt}$. Under capital market equilibrium, the unit housing price equals the annual rent divided by interest rate $\kappa_{t}$.

There is a national competitive market for construction materials, whereas land costs are city-specific, rising with local housing demand $\mathcal{H}_{mt}$. The marginal cost function can therefore be parametrized as the following form:
\begin{equation}
\text{MC}\left(\text{CC}_{mt},\text{LC}_{mt}\right)=\text{CC}_{mt}\cdot\mathcal{H}_{mt}^{\psi_{m}},\label{eq:housing03}
\end{equation}
where $\psi_{m}$ is the inverse elasticity of housing supply in city $m$. Following \cite{diamond2016determinants}, we further specify $\psi_{m}$ to be a function of the geographic and regulatory constraints imposed on city $m$:
\begin{equation}
\psi_{m}=\psi_{0}+\psi_{1}\cdot\exp\left(\chi_{m}^{\text{geo}}\right)+\psi_{2}\cdot\exp\left(\chi_{m}^{\text{reg}}\right),\label{eq:housing04}
\end{equation}
where $\chi_{m}^{\text{geo}}$ is a land unavailability index measuring the share of land in the city that is unavailable for housing development due to geographic constraints and was developed by \cite{saiz2010geographic}, while $\chi_{m}^{\text{reg}}$ is a land use regulation index measuring the stringency of local regulatory constraint on property development and was developed by \cite{gyourko2008new}. Both measures are used in \cite{saiz2010geographic} and have been widely adopted as determinants of the elasticity of housing supply.\footnote{$\left(\chi_{m}^{\text{geo}},\chi_{m}^{\text{reg}}\right)$ are taken as time-invariant in our model, since they exhibit small changes from 1980 to 2000.}

\subsection{Marriage Market}

Individuals from each city make decisions about whether to marry, whom to marry, and how much of both tradable goods and housing to consume. The utility maximization problem can be framed as a two-step procedure by backward induciton. First, for each given marital status, individuals optimize their consumption of tradable goods and housing to maximize utility. Then they choose their spouses or decide to remain single according the optimal utility in the first step.

To model marriage decisions, we adopt the framework of \cite{choo2006marries}, which assumes a frictionless matching market
with transferable utility. In this setting, individuals are categorized into discrete types, while each person has idiosyncratic preferences for spouses of different types. The concept of transferable utility implies that the total marital surplus generated by a couple is redistributed between the husband and wife through a marital transfer. This transfer, determined by the supply and demand of different types of individuals in a given marriage market, functions as the \textquotedblleft price\textquotedblright{} that regulates the market\textquoteright s equilibrium. 

In our framework, individuals are characterized by their education level, and each local marriage market consists of four sub-marriage markets: (1) the high-skill male, high-skill female market; (2) the high-skill male, low-skill female market; (3) the low-skill male, high-skill female market; (4) the low-skill male, low-skill female market. Each sub-marriage market has its own equilibrium transfer price.

\subsubsection*{Utility of Singles}

Single individuals maximize their utility by choosing an optimal consumption bundle of tradable goods and housing. The utility of a single individual $i$ living in city $m$ is given by: 
\begin{align*}
U_{imt}^{s}  =\max_{c_{imt},h_{imt}} \Big\{ \ln \big[ h_{imt}^{\zeta}c_{imt}^{1-\zeta}
 & \times\exp\left(f\left(\mathcal{A}_{mt};  g_i, e_i \right)+\sigma^{\epsilon}\epsilon_{it}^{s}+\sigma^{\nu}\nu_{imt}\right) \big] \Big\}
  \shortintertext{subject to} P_{t}c_{imt}+R_{mt}h_{imt}&\le W_{imt}
\end{align*}
Individuals derive pecuniary utility from the consumption of  tradable goods $c$ and housing $h$. In addition, they derive utility $f\left(\mathcal{A}_{mt};  g_i, e_i \right)$ from the amenity bundle $\mathcal{A}_{mt}$ of city $m$, and individuals of different education and gender may value amenity differently. $\epsilon_{it}^{s}$ reflects idiosyncratic preference for being single, while $\sigma^{\epsilon}$ captures its scale. Similarly, individuals also have idiosyncratic preference for each city $\sigma^{\nu}\nu_{imt}$.

Based on the above utility form, the indirect utility function for singles is given by 
\begin{align}
V_{imt}^{s} & =\ln W_{imt}-\zeta\ln R_{mt}+f\left(\mathcal{A}_{mt};  g_i, e_i \right)+\sigma^{\epsilon}\epsilon_{it}^{s}+\sigma^{\nu}\nu_{imt}  \label{eq:us} 
\end{align}

\subsubsection*{Utility of Married Individuals}

For married individuals, utility differs from singles in three aspects: (1) There exists nonpecuniary utility such as love and companionship;  (2) Pecuniary utility benefits from the household economy of scale; (3)Individual utility also depends on intrahousehold resource sharing and marital transfer.

Specifically, the utility of individual $i$ married to individual $j$, from the household k, is given as follows: 
\begin{align*}
U_{ikmt}^{e_{j}} & =\max_{\underset{h_{kmt}}{c_{kmt},}}\Big\{\ln\Big[\left(\frac{h_{kmt}^{\zeta}c_{kmt}^{1-\zeta}}{1+\chi}\right)\times\text{\ensuremath{\exp}}\left(\mu_{t}^{e_{j}}\left(g_i, e_{i}\right)+\tau_{mt}^{e_{j}}\left(g_{i},e_{i}\right)\right.\\
 & \ \ \ \ \ \ \ \ \ \ \ \ \ \ \left.+f\left(\mathcal{A}_{mt} ; g_{i},e_{i}\right)+\sigma^{\epsilon}\epsilon_{it}^{e_{j}}+\sigma^{\nu}\nu_{imt}\right)\Big]\Big\}\\
\text{subject to} & \ \ P_{t}c_{kmt}+R_{mt}h_{kmt}\le W_{imt}+W_{jmt}
\end{align*}
Following a unitary perspective, couples choose the tradable good and housing consumption bundle $\left(c_{mt},h_{mt}\right)$
jointly. There is household economy of scale. Each individual enjoys $1/\left(1+\chi\right)$ share of the consumption good, where $\chi\in[0,1]$ is the equivalence scale, capturing that a two-person household requires less than twice the consumption of a single-person household to achieve the same level of utility.

In addition, an individual $i$ married with individual $j$ additionally obtains marital utility:
\begin{align*}
\mu_{t}^{e_{j}}\left(g_{i},e_{i}\right)+\tau_{mt}^{e_{j}}\left(g_{i},e_{i}\right)+\sigma^{\epsilon}\epsilon_{it}^{e_{j}},
\end{align*}
where $\mu_{t}^{e_{j}}\left(g_{i},e_{i}\right)$ captures the  non-pecuniary preference of individual $i$ of type $(g_{i},e_{i})$ by marrying partner $j$ of type $(g_{j},e_{j})$ relative to being single. It is a reduced-form way to capture the attractiveness of a type of martial partners beyond financial factors. As we have no prior knowledge, we allow  it to be different by education, asymmetric across genders, and varying over time.\footnote{We assume that the nonpecuniary preferences do not vary by city to have enough degree of freedom and avoid overfitting.}

$\tau_{mt}^{e_{j}}\left(g_{i},e_{i}\right)$ captures the marital transfer that a type $\left(g_{i},e_{i}\right)$ person would receive if she is married to a type $\left(g_{j},e_{j}\right)$ spouse, satisfying
\begin{align*} \tau_{mt}^{e_{j}}\left(g_{i},e_{i}\right)=-\tau_{mt}^{e_{i}}\left(g_{j},e_{j}\right)
\end{align*}
This transfer mediates the demand and supply in each sub-marriage market indexed by $\left(e_{M},e_{F}\right)$ in each city. For instance, it may capture the allocation of private leisure after intrahousehold bargaining on chores. For a stable matching, low-skill wife may assume more domestic work and sacrifice private leisure in exchange of the market good consumption contributed by her high-skill husband who earns higher income. Marital transfer may also depend on local sex ratio. For instance, when high-skill males outnumber low-skill females in a market, the demand by low-skill females for high-skill males falls short of the demand by high-skill males for low-skill females. As a result, high-skill males tend to increase their transfers $-\tau_{mt}^{F,H}\left(L\right)$ to low-skill females. The marriage market clears when the demand by $e_{i}$-type male for $e_{j}$-type female is equal to the demand by $e_{j}$-type female for $e_{i}$-type male for all $e_{i},e_{j}\in\left\{ H,L\right\} $. Following \cite{choo2006marries}, we assume that each individual also obtains an idiosyncratic partner-type-specific utility $\epsilon_{it}^{e_{j}}$.

The indirect utility for individual $i$ married to spouse with education $e_{j}$ can therefore be written as: 
\begin{align}
\label{eq:um}
V_{imt}^{e_{j}} & =\ln\left(W_{imt}+W_{jmt} \right)-\zeta\ln R_{mt}-\ln\left(1+\chi\right)\nonumber \\
 & \ \ \ \ +\mu_{t}^{e_{j}}\left(g_{i},e_{i}\right)+\tau_{mt}^{e_{j}}\left(g_{i},e_{i}\right)+f\left(\mathcal{A}_{mt} ; g_{i},e_{i}\right)+\sigma^{\epsilon}\epsilon_{it}^{e_{j}}+\sigma^{\nu}\nu_{imt} 
\end{align}

Finally, let $p_{mt}^{e_{j}}\left(g_{i},e_{i}\right),p_{mt}^{g_{i},e_{i}}\left(s\right)$ denote the probabilities of individual $i$ of education type $e_i$ matching with a spouse of education type $e_{j}$ or remaining single. Assume $\epsilon_{it}^{s},\epsilon_{it}^{H},\epsilon_{it}^{L}$ as independently and identically distributed, following the distribution
of $\text{Gumbel}\left(-\gamma,1\right)$.\footnote{$\gamma$ is the Euler constant.} Then we obtain the closed form expression for the conditional choice probability of marriage:
\begin{align*}
p_{mt}^{e_{j}}\left(g_{i},e_{i}\right) & =\frac{\exp\left(\overline{V}_{mt}^{e_{j}}\left(g_{i},e_{i}\right)\right)}{\exp\left(\overline{V}_{mt}^{H}\left(g_{i},e_{i}\right)\right)+\exp\left(\overline{V}_{mt}^{L}\left(g_{i},e_{i}\right)\right)+\exp\left(\overline{V}_{mt}^{s}\left(g_{i},e_{i}\right)\right)}\\
p_{mt}^{s}\left(g_{i},e_{i}\right) & =1-p_{mt}^{H}\left(g_{i},e_{i}\right)-p_{mt}^{L}\left(g_{i},e_{i}\right)
\end{align*}
where $\overline{V}_{mt}^{e_j}\left(g_{i},e_{i}\right)  = \frac{1}{\sigma^{\epsilon}}\left[\ln\left(W_{imt}+W_{jmt}\right)-\zeta\ln R_{mt}-\ln\left(1+\chi\right)\right. \left.+\mu_{t}^{e_{j}}\left(g_{i},e_{i}\right)+\tau_{mt}^{e_{j}}\left(g_{i},e_{i}\right)\right] $
and \\
$\overline{V}_{mt}^{s}\left(g_{i},e_{i}\right)  = \frac{1}{\sigma^{\epsilon}}\left[\ln\left(W_{imt}\right)-\zeta\ln R_{mt} \right] $.\footnote{ Note that individual's wage $W_{imt}$ differs by their education, gender, and marital status. Therefore, the subscript $i$ in $\overline{V}_{imt}^{g_{i},e_{i}}$ can be omitted for each given marriage choice.}

\subsection{Location Choice}

Location choice is based on the expected utility of living in each city. When making location decisions, we assume individuals have rational expectations over equilibrium populations, wages, rents, and marriage transfers. The value of city $m$ to individual $i$, prior to the realization of marital preference shocks, is given by
\begin{align}
\Psi_{imt}\left(g_{i},e_{i}\right)=\mathbb{E}_{\varepsilon}\left[\max\left\{ V_{imt}^{s},V_{imt}^{H},V_{imt}^{L}\right\} \right]=\mathcal{V}_{mt}\left(g_{i},e_{i}\right)+f\left(\mathcal{A}_{mt};g_{i},e_{i}\right)+\sigma^{\nu}\nu_{imt},\label{eq:loc00}
  \shortintertext{where}
\mathcal{V}_{mt}\left(g_{i},e_{i}\right)=\sigma^{\epsilon}\ln\left[\exp\left(\overline{V}_{mt}^{s}\left(g_{i},e_{i}\right)\right)+\exp\left(\overline{V}_{mt}^{H}\left(g_{i},e_{i}\right)\right)+\exp\left(\overline{V}_{mt}^{L}\left(g_{i},e_{i}\right)\right)\right] \notag
\end{align}
As mentioned earlier, $f\left(\mathcal{A}_{mt}; g_{i}, e_{i} \right)$ represents the utility that individual $i$ derives from the amenity bundle $\mathcal{A}_{mt}$ offered by city $m$. We assume $\mathcal{A}_{mt}$ is exogenous and
\begin{equation}
f\left(\mathcal{A}_{mt}; g_i, e_i\right)=\eta\mathcal{A}_{mt}^{o}+a_{mt}\left(g_{i},e_{i}\right),\label{eq:loc02}
\end{equation}
where $\mathcal{A}_{mt}^{o}$ is an amenity index developed by \cite{diamond2016determinants} that measures a city's observed retail environment, transportation infrastructure, crime rate, environmental quality, school quality, and job quality, while $a_{mt}\left(g_{i},e_{i}\right)$ represents (the valuation of) unobserved amenities. 

Let $\Omega_{mt}\left(g_{i},e_{i}\right)=\mathcal{V}_{mt}\left(g_{i},e_{i}\right)-\sigma^{\epsilon}\overline{V}_{mt}^{s}\left(g_{i},e_{i}\right)$
be the \emph{marital surplus} of city $m$ -- the expected value of getting married in the city in excess of the value of being single. Then \eqref{eq:loc00} can be written as
\begin{equation}
\Psi_{imt}\left(g_{i},e_{i}\right)=\ln W_{mt}-\zeta\ln R_{mt}+\Omega_{mt}\left(g_{i},e_{i}\right)+f\left(\mathcal{A}_{mt}; g_i, e_i\right)+\sigma^{\nu}\nu_{imt},\label{eq:loc01}
\end{equation}
i.e., individuals choose where to live based on each city's wage, rent, marital prospect, amenities, as well as their own idiosyncratic tastes. 

Let $q_{t}^{m}\left(g_{i},e_{i}\right)$ denote the probability of individual $i$ choosing location $m$. Then assuming $\nu_{imt}\overset{\text{i.i.d.}}{\sim}\text{Gumbel}\left(-\gamma,1\right)$, we can derive the closed-form expression for location choice probabilities:
\begin{equation}
q_{t}^{m}\left(g_{i},e_{i}\right)=\frac{\exp\left(\overline{\Psi}_{mt}\left(g_{i},e_{i}\right)\right)}{\sum_{m^{'}=1}^{M}\exp\left(\overline{\Psi}_{m^{'}t}\left(g_{i},e_{i}\right)\right)},\label{eq:locationchoice}
\end{equation}
where $\overline{\Psi}_{mt}\left(g_{i},e_{i}\right)=\frac{1}{\sigma^{\nu}}\left(\ln W_{mt}-\zeta\cdot\ln R_{mt}+\Omega_{mt}\left(g_{i},e_{i}\right)+f\left(\mathcal{A}_{mt}; g_i, e_i\right) \right).$

\subsection{Equilibrium}

In equilibrium, each city's local labor markets, housing market, and marriage markets clear. Individuals' location choices determine the labor supply, the housing demand, as well as both the supply and demand of marriage. 

Specifically, equilibrium of each city is characterized by local wage rates, rents and marital transfers between each type of couples $\left(w_{mt}^{H*},w_{mt}^{L*},R_{mt}^{*},\tau_{mt}^{H}(MH)^{*},\tau_{mt}^{L}(MH)^{*},\tau_{mt}^{H}(ML)^{*},\tau_{mt}^{L}(ML)^{*}\right)$, as well as by local population $\left(N_{mt}^{MH*},N_{mt}^{ML*},N_{mt}^{FH*},N_{mt}^{FL*}\right)$ such that: 
\begin{enumerate}
\item The demand for high-skilled workers equals their supply, measured in
effective units. 
\begin{align*}
\mathcal{L}_{mt}^{H*}= & \sum_{i} q_{t}^{m}\left(g_{i},e_{i}\right) \cdot\ell_{i}^{g_{i},H}   \text{ \qquad and \qquad }
w_{mt}^{H*}=\left.\partial Y_{mt}\right/\partial\mathcal{L}_{mt}^{H}
\end{align*}
\item The demand for low-skilled workers equals the supply, measured in
the effect unit. 
\begin{align*}
\mathcal{L}_{mt}^{L*}= & \sum_{i} q_{t}^{m}\left(g_{i},e_{i}\right) \cdot\ell_{i}^{g_{i},L}   \text{ \qquad and \qquad }
w_{mt}^{L*}=\left.\partial Y_{mt}\right/\partial\mathcal{L}_{mt}^{L}
\end{align*}
\item The demand and supply of housing are equal. 
\begin{align*}
\ln R_{mt}^{*} & =\ln\kappa_{t}+\ln\text{CC}_{mt}+\psi_{m}\cdot\ln\mathcal{H}_{mt}^{*}\\
\mathcal{H}_{mt}^{*} & =\frac{\zeta}{R_{mt}^{*}}\mathcal{W}_{mt}
\end{align*}
\item The demand and supply of each type of marriage candidates are equal. 
\begin{align*}
p_{mt}^{H}\left(M,H\right)\cdot N_{mt}^{MH*}= & \ p_{mt}^{H}\left(F,H\right)\cdot N_{mt}^{FH*}\\
p_{mt}^{L}\left(M,H\right)\cdot N_{mt}^{MH*}= & \ p_{mt}^{H}\left(F, L\right)\cdot N_{mt}^{FL*}\\
p_{mt}^{H}\left(M,L\right)\cdot N_{mt}^{ML*}= & \ p_{mt}^{L}\left(F, H\right)\cdot N_{mt}^{FH*}\\
p_{mt}^{L}\left(M, L\right)\cdot N_{mt}^{ML*}= & \ p_{mt}^{L}\left(F, L\right)\cdot N_{mt}^{FL*}
\end{align*}
\end{enumerate}

\section{Empirical Implementation}
\label{section:ei}

\subsection{Data}
We estimate the structural model using the 5 percent sample of the US Census for 1980, 1990, and 2000, obtained from the Integrated Public Use Microdata Series (IPUMS). These data provide individual-level observations on a range of demographic and economic characteristics, including education, earnings, marital status, and geographic location. Additionally, unique family identifiers enable the matching of spouses within households.

Our analysis focuses on civilians aged 25 to 55 who work full time (35 hours or more per week), with the exception of married women, who are included regardless of their employment status. The sample of married households is restricted to heterosexual couples, thereby excluding same-sex marriages from the analysis. Individuals are indexed by gender and educational attainment: those with at least four years of college are classified as high-skill, while all others are classified as low-skill.  To ensure comparability over time, all wage and price data are adjusted for inflation using the Consumer Price Index (CPI) and expressed in constant 1999 dollars.

The metropolitan statistical area (MSA) serves as the primary geographic unit for our analysis. Our dataset includes 218 MSAs that are consistently defined across the three decades under study, enabling the quantitative model to accurately capture the spatial distributions of skills and marital outcomes. A significant advantage of using Census data is its large sample size, which provides sufficient observations within each MSA, allowing for reliable calculation of marriage and labor market outcomes at the granular city level.

We augment the Census data by incorporating additional measures on key city characteristics. To quantify differences in city attractiveness, we use the urban amenity index developed by \citet{diamond2016determinants}, which aggregates data from 15 amenity indicators---such as transportation, crime rates, environmental quality, and school quality---into a single index based on principal component analysis. To model housing supply elasticity, we use the land unavailability index from \citet{saiz2010geographic}, which captures geographic constraints on housing development, as well as the Wharton land use regulation index by \citet{gyourko2008new}, which measures the stringency of local regulatory constraints on property development.

\subsection{Estimation on Marriage}

The objective of the marriage estimation is to uncover the nonpecuniary preference $\mu_t^{e_j}\left(g_{i},e_{i}\right)$ and the scale parameter of the idiosyncratic marital preference $\sigma^{\epsilon}$. However, a great challenge in identifying these parameters lies in the fact that marital transfers $\tau_{mt}^{e_j}(g_i,e_i)$ are unobserved. We address this challenge by exploiting the rich variation in marriage outcomes across cities to obtain a joint identification over $\mu_t^{e_j}\left(g_{i},e_{i}\right)$, $\sigma^{\epsilon}$ and $\tau_{mt}^{e_j}\left(g_{i},e_{i}\right)$.

Specifically, by taking the difference between the indirect utility functions defined in Equation~\eqref{eq:us} and \eqref{eq:um}, we derive the marital choice probability equation as follows: 
\begin{align}
\ln p_{mt}^{e_{j}}\left(g_{i},e_{i}\right)- \ln p_{mt}^{s}\left(g_{i},e_{i}\right) & =\overline{V}_{mt}^{e_{j}}\left(g_{i},e_{i}\right)-\overline{V}_{mt}^{s}\left(g_{i},e_{i}\right)\label{eq:mar0}\\
 & =\widetilde{\mu}_{t}^{e_{j}}\left(g_{i},e_{i}\right)+\widetilde{\tau}_{mt}^{e_{j}}\left(g_{i},e_{i}\right)+\frac{1}{\sigma^{\epsilon}}\ln\left(\frac{w_{imt}+w_{jmt}}{w_{imt}\left(1+\chi\right)}\right),\nonumber \\
  \shortintertext{where}
\widetilde{\mu}_{t}^{e_{j}}\left(g_{i},e_{i}\right)\equiv & \left(\sigma^{\epsilon}\right)^{-1}\mu_{t}^{e_{j}}\left(g_{i},e_{i}\right)\text{ and }\widetilde{\tau}_{mt}^{e_{j}}\left(g_{i},e_{i}\right)\equiv\left(\sigma^{\epsilon}\right)^{-1}\tau_{mt}^{e_{j}}\left(g_{i},e_{i}\right).\nonumber 
\end{align}
Equation~\eqref{eq:mar0} represents an economic model. To link it to the data, we replace the probabilities $p_{mt}^{e_{j}}\left(g_{i},e_{i}\right)$ with their sample estiamtes $\hat{p}_{mt}^{e_{j}}\left(g_{i},e_{i}\right)$, yielding an empirical model. This model is  estimated by exploiting geographic and temporal variation in marital choice probabilities using the generalized method of moments (GMM). While the $\tau's$ are endogenous variables determined by local marriage market equilibria, we treat their counterparts in reality as unknown parameters to be recovered.

It is important to emphasize that the framework proposed by \cite{choo2006marries} is only just-identified: the marital choice probabilities on the left-hand side, for individuals of type $e_i$, identify only the total marital surplus of marrying a spouse of type $e_j$ relative to remaining single. Consequently, in a single national marriage market, the structural parameters $\widetilde{\mu}_{t}^{e_{j}}\left(g_{i},e_{i}\right)$ and $\widetilde{\tau}_{mt}^{e_{j}}\left(g_{i},e_{i}\right)$ cannot be separately identified. However, our framework of local marriage markets, which leverages the rich variation in marital outcomes across cities, allows for the identification of these structural objects. This approach uncovers novel insights related to nonpecuniary utility $\mu$ and the transfers $\tau$. For the remaining parameters, the equivalence scale $\chi$ is calibrated  to 0.7, following \cite{greenwood2016technology} and \cite{organisation2013oecd}. $\phi_t^H \text{ and } \phi_t^L$ are determined by observed wage gaps adjusted for effective labor, while $\delta_{mt}^H \text{ and } \delta_{mt}^L$ are calibrated using the labor force participation rates of married females.

The basic intuition for identification is as follows. The model is over-identified, as Equation \eqref{eq:mar0} involves more data points of marital choices than objects to be estimated.\footnote{There are $N^{g}\cdot N^{e_i}\cdot N^{e_j}\cdot N^{m}\cdot N^{t}=5,208$ data points of marital choice probabilities. The unknown marital transfers account for  $N^{e_i} \cdot N^{e_j} \cdot N^{m}\cdot N^{t}=2,604$ objects, while $\mu$ involes $N^{g}\cdot N^{e_i}\cdot N^{e_j}=8$ parameters, with an additional $\sigma^\epsilon$.  $N^{g}, N^{e_i}, N^{e_j}, N^{m}, N^{t}$ represent the dimensions of gender, own education,  partner's education, location, and time, respectively.} The pecuniary attraction of a potential partner $j$ depends on their wage but also varies with the own wage of individual $i$. Variation in marital choices and wages across individuals, cities, and time contributes to the identification of $\sigma^{\epsilon}$. The nonpecuniary utility $\mu$ captures the attraction of a partner of type $e_{j}$ to an individual $i$ of gender $g_{i}$ and type $e_{i}$, beyond monetary benefits. It is identified by marital choice probabilities for partners of various education levels, net of the pecuniary effects of education.  Martial transfers $\tau$ are mainly identified by local marriage market equilibria. For instance, in a city with a significantly larger supply of high-skill males but where the PMH of high-skill females is similar to other cities, this pattern is rationalized by higher martial transfers received by hill-skill females in that city from high-skill males. The identification of $\tau$ is further sharpened by the constraint $\tau_{mt}^{e_{j}}\left(g_{i},e_{i}\right)=-\tau_{mt}^{e_{i}}\left(g_{j},e_{j}\right)$, which ensures symmetry in transfers.

\subsection{Estimation on Labor and Housing Markets}

\subsubsection{Bartik Labor Demand Shocks}

Our estimation for labor demand, housing supply, and labor supply follows \citet{diamond2016determinants} and employ Bartik labor demand shocks as shift-share instrumental variables. These shocks capture exogenous shifts in local labor demand, driven by national productivity changes across industries that  affect cities differentially based on their industrial composition. Specifically, let $k$ denote industry, $t_{0}$ represent the base year 1980, and $\Delta x_{t}=x_{t}-x_{t_{0}}$ indicate the change in a variable $x$ relative to its level in 1980. For skill levels $e\in\left\{ H,L\right\} $, the Bartik labor demand shock is constructed as follows:
\begin{equation*}
\Delta B_{mt}^{e}=\frac{1}{\mathcal{L}_{mt_{0}}^{e}}\sum_{k}\mathcal{L}_{mkt_{0}}^{e}\Delta\ln w_{-mkt}^{e},
\end{equation*}
where $\mathcal{L}_{mkt_{0}}^{e}$ denotes the total employment in industry $k$ for skill level $e$ in city $m$ during the base year, and $\ln w_{-mkt}^{e}$ represents the national log wage growth for skill level $e$ in industry $k$ at time $t$, excluding city $m$. The leave-one-out estimator ensures that local wage changes in city $m$ do not bias the measurement of its productivity shocks. By interacting a city\textquoteright s industrial composition with national wage growth in those industries, the Bartik shocks capture local labor demand variation that is plausibly exogenous to city-specific conditions.

\subsubsection{Labor Demand}

From the production function \eqref{eq:prod00}, we can derive the following equation:
\begin{equation}
\ln w_{mt}^{e}=\ln A_{mt}+\left(\frac{1}{\rho}-1\right)\ln G_{mt}+\left(\rho-1\right)\ln\mathcal{L}_{mt}^{e}+\ln\alpha_{mt}^{e},\label{eq:lde01}
\end{equation}
where $G_{mt}=\left[\alpha_{mt}\left(\mathcal{L}_{mt}^{H}\right)^{\rho}+\left(1-\alpha_{mt}\right)\left(\mathcal{L}_{mt}^{L}\right)^{\rho}\right]$,
$\alpha_{mt}^{H}=\alpha_{mt}$, and $\alpha_{mt}^{L}=1-\alpha_{mt}$.
Taking the difference of high-skill relative to low-skill wages, and then further differencing with respect to 1980, we obtain
\begin{equation}
\Delta\ln\frac{w_{mt}^{H}}{w_{mt}^{L}}=\left(\rho-1\right)\Delta\ln\frac{\mathcal{L}_{mt}^{H}}{\mathcal{L}_{mt}^{L}}+\Delta\epsilon_{mt},\label{eq:lde02}
\end{equation}
where $\Delta\epsilon_{mt}=\Delta\ln\alpha_{mt}^{H}-\Delta\ln\alpha_{mt}^{L}$ represents exogenous shifts in the demand for high-skill labor relative to low-skill labor. We can further decompose $\Delta\epsilon_{mt}$ into 
\begin{equation}
\Delta\epsilon_{mt}=\gamma_{mt}^{1}\Delta B_{mt}^{H}+\gamma_{mt}^{2}\Delta B_{mt}^{L}+\Delta\widetilde{\epsilon}_{mt},\label{eq:lde03}
\end{equation}
where $\Delta\widetilde{\epsilon}_{mt}^{e}$ captures the part of the local exogenous productivity change that is uncorrelated with
the Bartik labor demand shocks.

Substituting \eqref{eq:lde03} into \eqref{eq:lde02}, we arrive at the labor demand estimating equation
\begin{equation}
\Delta\ln\frac{w_{mt}^{H}}{w_{mt}^{L}}=\left(\rho-1\right)\Delta\ln\frac{\mathcal{L}_{mt}^{H}}{\mathcal{L}_{mt}^{L}}+\gamma_{mt}^{1}\Delta B_{mt}^{H}+\gamma_{mt}^{2}\Delta B_{mt}^{L}+\Delta\widetilde{\epsilon}_{mt}\label{eq:lde04}
\end{equation}

To estimate $\rho$, we follow \cite{diamond2016determinants} and construct instruments for $\Delta\ln\left(\left.\mathcal{L}_{mt}^{H}\right/\mathcal{L}_{mt}^{L}\right)$ by interacting local land characteristics with the Bartik labor demand shocks. Let 
\[
\Delta Z_{mt}=\left(\Delta B_{mt}^{H}\chi_{m}^{\text{geo}},\Delta B_{mt}^{H}\chi_{m}^{\text{geo}},\Delta B_{mt}^{L}\chi_{m}^{\text{reg}},\Delta B_{mt}^{L}\chi_{m}^{\text{reg}}\right)
\]
be our set of instruments. Intuitively, in response to rising wages driven by local labor demand shocks, local labor supply responses vary due to constraints imposed by land and housing regulations. Cities with more elastic housing supply tend to experience larger inflows
of labor. These resulting labor supply shocks contribute to $\Delta\ln\left(\left.\mathcal{L}_{mt}^{H}\right/\mathcal{L}_{mt}^{L}\right)$
but are plausibly uncorrelated with $\Delta\widetilde{\epsilon}_{mt}^{e}$. 

After estimating equation \eqref{eq:lde04} and obtaining $\widehat{\rho}$, we can back out the values of the other technology parameters as 
\begin{align*}
\widehat{\alpha}_{mt} & =\frac{w_{mt}^{H}\left(\mathcal{L}_{mt}^{H}\right)^{1-\widehat{\rho}}}{w_{mt}^{H}\left(\mathcal{L}_{mt}^{H}\right)^{1-\widehat{\rho}}+w_{mt}^{L}\left(\mathcal{L}_{mt}^{L}\right)^{1-\widehat{\rho}}}\\
\widehat{A}_{mt} & =\frac{\left[w_{mt}^{H}\left(\mathcal{L}_{mt}^{H}\right)^{1-\widehat{\rho}}+w_{mt}^{L}\left(\mathcal{L}_{mt}^{L}\right)^{1-\widehat{\rho}}\right]^{\frac{1}{\widehat{\rho}}}}{\left[w_{mt}^{H}\mathcal{L}_{mt}^{H}+w_{mt}^{L}\mathcal{L}_{mt}^{L}\right]^{\frac{1}{\widehat{\rho}}-1}}
\end{align*}

\subsubsection{Housing Supply}

Substituting \eqref{eq:housing03} and \eqref{eq:housing04} into the the housing supply equation \eqref{eq:housing02}, and taking
difference with respect to 1980, we obtain a housing supply estimating equation
\begin{equation}
\Delta\ln R_{mt}=\Delta\ln\kappa_{t}+\left(\psi_{0}+\psi_{1}\cdot\exp\left(\chi_{m}^{\text{geo}}\right)+\psi_{2}\cdot\exp\left(\chi_{m}^{\text{reg}}\right)\right)\cdot\Delta\ln\mathcal{H}_{mt}+\Delta\varepsilon_{mt},\label{eq:hse02}
\end{equation}
where $\Delta\varepsilon_{mt}=\Delta\ln\text{CC}_{mt}$. We observe local rent $R_{mt}$ and land characteristics $\left(\chi_{m}^{\text{geo}},\chi_{m}^{\text{rent}}\right)$. From the housing demand equation \eqref{eq:housing00}, we obtain the equilibrium quantity of housing $\mathcal{H}_{mt}=\frac{\zeta}{R_{mt}}\mathcal{W}_{mt}$. Following \citet{moretti2013real}, we set the local good expenditure share $\zeta$ to be $0.62$, so that $\mathcal{H}_{mt}$ can be computed directly from data. Finally, to estimate equation \eqref{eq:hse02} and obtain estimates of the inverse housing supply elasticity parameters $\left(\psi_{0},\psi_{1},\psi_{2}\right)$, we again follow \citet{diamond2016determinants} and use
\[
\Delta Z_{mt}=\left(\Delta B_{mt}^{H},\Delta B_{mt}^{L},\Delta B_{mt}^{H}\chi_{m}^{\text{geo}},\Delta B_{mt}^{H}\chi_{m}^{\text{geo}},\Delta B_{mt}^{L}\chi_{m}^{\text{reg}},\Delta B_{mt}^{L}\chi_{m}^{\text{reg}}\right)
\]
as a set of instruments for $\Delta\ln\mathcal{H}_{mt}$. These instruments represent exogenous housing demand shocks, driven by the Bartik labor
demand shocks and their interaction with local land characteristics, and are plausibly orthogonal to unobserved changes in local construction costs $\Delta\ln\text{CC}_{mt}$.

\subsubsection{Location Choice}

The labor supply in each city is governed by individuals' location choices, which take into account each city's wage, rent, amenity, and marital surplus. The estimation of location choice preferences proceeds in two steps. First, given a reference city $m_{0}$, let
$u_{mt}\left(g_{i},e_{i}\right)=\overline{\Psi}_{mt}\left(g_{i},e_{i}\right)-\overline{\Psi}_{m_{0}t}\left(g_{i},e_{i}\right)$ be the scaled value of city $m$ relative to city $m_{0}$, net of idiosyncratic preferences.\footnote{In our estimation, we choose State College, PA as the reference city.} From Equation \eqref{eq:locationchoice}, we have
\begin{equation}
u_{mt}\left(g_{i},e_{i}\right)=\log q_{t}^{m}\left(g_{i},e_{i}\right)-q_{t}^{m_{0}}\left(g_{i},e_{i}\right)\label{eq:lce01}
\end{equation}
By plugging the observed location choice probabilities into \eqref{eq:lce01}, we can obtain estimates of $u_{mt}\left(g_{i}, e_{i}\right)$. In the second step, we express $u_{mt}\left(g_{i}, e_{i}\right)$ as 
\begin{equation}
u_{mt}\left(g_{i},e_{i}\right)=\varsigma_{t}(g_{i}, e_{i})+\beta_{1}\mathcal{V}_{mt}(g_{i}, e_{i})+\beta_{2}\mathcal{A}_{mt}^{o}+\xi_{mt}(g_{i}, e_{i}),\label{eq:lce02}
\end{equation}
where $\varsigma_{t}(g_{i}, e_{i})=\overline{\Psi}_{m_{0}t}(g_{i}, e_{i})$, $\beta_{1}=\left(\sigma^{\nu}\right)^{-1}$, $\beta_{2}=\left(\sigma^{\nu}\right)^{-1}\eta$, and $\xi_{mt}(g_{i}, e_{i})=\left(\sigma^{\nu}\right)^{-1}a_{mt}(g_{i}, e_{i})$. $\mathcal{A}_{mt}^{o}$ is observed, while $\mathcal{V}_{mt}(g_{i}, e_{i})$---the average value of living in city $m$ at time $t$ net of amenity
preferences---can be computed based on results from marriage estimation. Differencing \eqref{eq:lce02} with respect to 1980 gives
\begin{equation}
\Delta u_{mt}(g_{i}, e_{i})=\Delta\varsigma_{t}(g_{i}, e_{i})+\beta_{1}\Delta\mathcal{V}_{mt}(g_{i}, e_{i})+\beta_{2}\Delta\mathcal{A}_{mt}^{o}+\Delta\xi_{mt}(g_{i}, e_{i})\label{eq:lce03}
\end{equation}

Equation \eqref{eq:lce03} serves as our labor supply estimating equation, where $\Delta\varsigma_{t}(g_{i}, e_{i})+\mathbb{E}\left[\Delta\xi_{mt}(g_{i}, e_{i})\right]$ can be considered as $\left(g,e\right)-$specific fixed effects. Given the assumption of exogenous amenities, we can estimate the equation by using the Bartik labor demand shocks 
\[
\Delta Z_{mt}=\left(\Delta B_{mt}^{H},\Delta B_{mt}^{L}\right)
\]
as the instruments for $\Delta\mathcal{V}_{mt}^{g_{i}e_{i}}$. Then, given the estimated $\left(\widehat{\beta}_{1},\widehat{\beta}_{2}\right)$,
we can reconstruct the model parameters as $\widehat{\sigma}^{\nu}=\left.1\right/\widehat{\beta}_{1},\widehat{\eta}=\left.\widehat{\beta}_{2}\right/\widehat{\beta}_{1}$. 

\section{Estimation Results and Model Fits}

\subsection{Parameter Estimates}

% Figure
\begin{figure}[t]
\centering 
\begin{minipage}[c]{0.99\textwidth} 
\centering
\includegraphics[width=0.7\columnwidth]{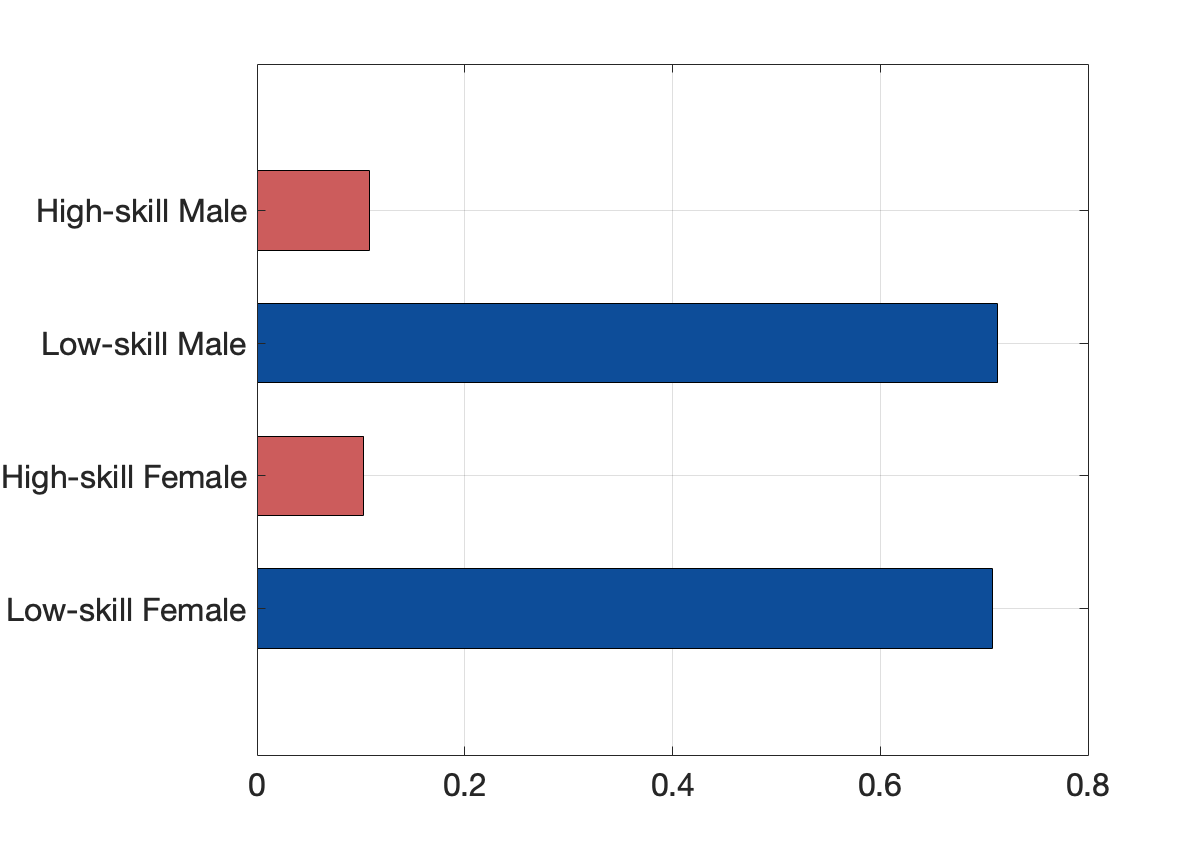} 
\caption{Difference in Nonpecuniary Utility: Low-Skill vs. High-Skill Spouses}
\label{fig:mu} 

\vspace{1.0em} 
\raggedright
\scriptsize 
\noindent\parbox{1.0\textwidth}{
    \textit{Notes:} This figure illustrates the average difference in the nonpecuniary utility of marrying a low-skill spouse relative to that of marrying a high-skill spouse, calculated for each individual type.
    }
\end{minipage}
\end{figure}
% end 

% Table
\begin{table}[t] 
\centering
\footnotesize 
\caption{Parameter Estimates: Labor Demand, Housing Supply, \\Marriage, and Location Choice}
\label{tab:Parameter-Estimates}
\begin{minipage}{0.99\textwidth} 
\centering 
\tabcolsep=6pt 
\begin{tabular}{p{0.35\textwidth}>{\centering}p{0.1\textwidth}p{0.35\textwidth}>{\centering\arraybackslash}p{0.1\textwidth}}
\toprule 
\toprule
\multicolumn{2}{l}{\textit{A. Labor Demand}} & \multicolumn{2}{l}{\textit{C. Marriage Choice}} \tabularnewline
\cmidrule(lr){1-2} \cmidrule(lr){3-4}
{\footnotesize substitution parameter ($\rho$)} & {\footnotesize 0.577} & {\footnotesize scale of idiosyncratic shocks ($\sigma^{\epsilon}$)} & {\footnotesize 2.728} \tabularnewline
 & {\footnotesize (0.160)} & & {\footnotesize (0.000)} \tabularnewline
\midrule
\multicolumn{2}{l}{\textit{B. Housing Supply}} & \multicolumn{2}{l}{\textit{D. Location Choice}} \tabularnewline
\cmidrule(lr){1-2} \cmidrule(lr){3-4}
{\footnotesize base elasticity ($\psi_{0}$)} & {\footnotesize -0.001} & {\footnotesize scale of idiosyncratic shocks ($\sigma^{\nu}$)} & {\footnotesize 7.072} \tabularnewline
 & {\footnotesize (0.102)} & & {\footnotesize (0.001)} \tabularnewline
{\footnotesize land unavailability ($\psi_{1}$)} & {\footnotesize 0.015} & {\footnotesize amenity preference} ($\eta$) & {\footnotesize 0.085} \tabularnewline
 & {\footnotesize (0.009)} & & {\footnotesize (0.002)} \tabularnewline
{\footnotesize land regulation ($\psi_{2}$)} & {\footnotesize 0.051} & & \tabularnewline
 & {\footnotesize (0.020)} & & \tabularnewline
\bottomrule 
\end{tabular}

\vspace{0.5em} 
\raggedright
\scriptsize 
\noindent\parbox{1.0\textwidth}{
    \textit{Notes:} Standard errors are shown in parentheses. 
    }
\end{minipage}
\end{table}
% end 

Table \ref{tab:est_m} presents our estimates of nonpecuniary marital utilities for each type of individual over time. Since the utility of being single is normalized to zero, these estimates reflect the utility of marrying different types of spouses relative to remaining single. On average, the nonpecuniary utility of marrying a low-skill spouse exceeds that of marrying a high-skill partner across all individual types, indicating that the observed prevalence of individuals married to low-skill spouses cannot be explained solely by monetary factors.

By individual type, Figure \ref{fig:mu} plots the difference in nonpecuniary utility between marrying a low-skill spouse and a high-skill spouse. The figure reveals a clear pattern of assortative nonpecuniary preferences, with high-skill individuals exhibiting a significantly stronger preference for high-skill partners compared to low-skill individuals. 

We also uncover novel patterns in marital transfers across cities. For instance, in high-skill cities, the large concentration of college-educated individuals tends to reduce the marital transfers they receive. Conversely, the higher wages in these cities increase their attractiveness in the marriage market, exerting an opposing effect by raising the transfers they receive. Since marital transfers are endogenous variables in our model and carry important welfare implications, we defer a detailed discussion to Section \ref{section:wf} on welfare analysis.

Table \ref{tab:Parameter-Estimates} reports the estimates for the remaining parameters in our model. In Panel A, the estimated elasticity of labor substitution is $\sigma=1\left/\left(1-\rho\right)\right.=2.36$, implying gross-substitutes between high-skill and low-skill labor.\footnote{Our estimate of the elasticity of substitution between high-skill and low-skill labor is in line with the literature. See \citet{katz_changes_1999} for a review.} Panel B presents the estimated inverse housing supply elasticity. Consistent with the literature, we find that housing supply is more inelastic in areas with less land availability and more stringent regulation. The magnitude of the inverse elasticity estimates is close to those of \citet{diamond2016determinants}, suggesting similar variation in housing supply elasticities across cities.

Panel C and Panel D report the estimated scale parameters from the marriage and location choice models, respectively. These parameters determine how responsive individuals are to the gains from marriage and to the value of a specific location. A  smaller $\sigma^{\epsilon}$ implies that individuals are more likely to marry a given type of spouse as the corresponding benefit increases. Similarly a smaller $\sigma^{\nu}$ suggests greater responsiveness to changes in location value, which, given our decadal data, implies a higher long-run migration elasticity. Notably, our estimated $\sigma^{\nu}$ match closely that of \citet{diamond2016determinants}, even though we assume exogenous local amenities.

\subsection{Model Fit}

% Figure
\begin{figure}[t]
\centering 
\begin{minipage}[c]{0.99\textwidth} 
%\centering
\includegraphics[height=7.8cm]{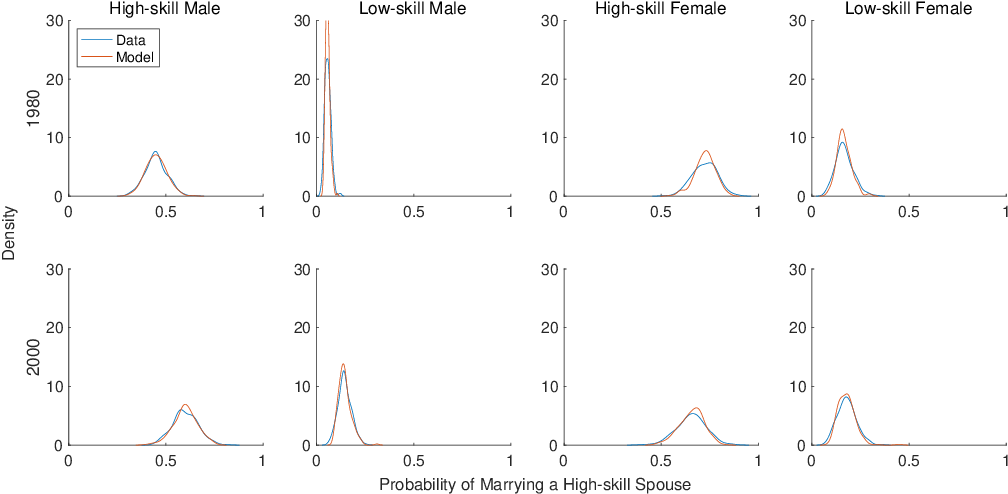} 
\vspace{0.5em} 
\caption{Model Fit: Marriage Choices over Partners, Levels}
\label{fig:Model-Fit-PMH-1}

\vspace{1.0em} 
\raggedright
\scriptsize 
\noindent\parbox{1.0\textwidth}{
    \textit{Notes:} This figure illustrates the distribution of the probability of marrying a high-skill spouse (PMH) across cities, differentiated by individual type, and compares observed data with predictions from the model.
    }
\end{minipage}
\end{figure}
% end

% Figure
\begin{figure}[t]
\centering 
\begin{minipage}[c]{0.99\textwidth} 
%\centering
\includegraphics[height=4cm]{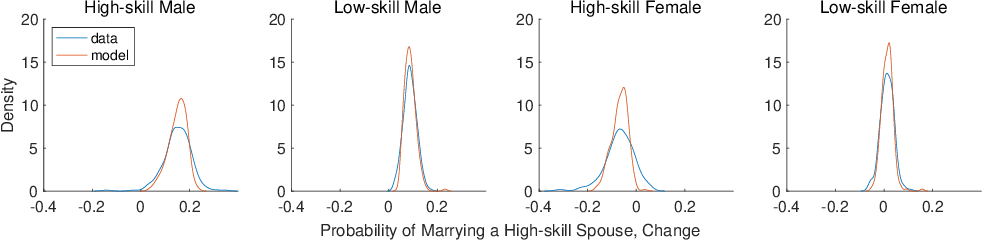}  
\caption{Model Fit: Marriage Choices over Partners, Changes}
\label{fig:Model-Fit-1-2}

\vspace{1.0em} 
\raggedright
\scriptsize 
\noindent\parbox{1.0\textwidth}{
    \textit{Notes:} This figure illustrates the distribution of the change in probability of marrying a high-skill spouse (PMH) across cities from 1980 to 2000, differentiated by individual type, and compares observed data with predictions from the model.
    }
\end{minipage}
\end{figure}
% end

To evaluate the performance of our model, we use the estimated model to predict marriage and location choices and compare the prediction with the observed data. For marital choices, we focus primarily on the fit of the key empirical outcome of this study---the choice of a spouse with a given education level, conditional on getting married. %The fit for the decision of whether to marry is provided in Appendix \ref{ap:fit}.

Figure \ref{fig:Model-Fit-PMH-1} presents the
in-sample fit for the PMH, with each plot showing the distribution of PMH across cities. Our estimated model is able to closely predict the cross-sectional spatial patterns of marriage, both by individual type and across time periods. While this figure shows the model fit for spatial distributions,  Figure \ref{fig:Model-Fit-PMH-2} provides additional insights by showing a scatter plot of the observed data against the model's predictions for each city.

Different cities also have experienced different evolutions in local wages, education distributions, as well as marital outcomes. Figure \ref{fig:Model-Fit-1-2} shows the spatial distributions for the changes in the PMH from 1980 to 2000. Our model is also able to capture these changes across cities reasonably well.

Figure \ref{fig:Model-Fit-2} presents the model's fit for location choices by comparing the observed and the predicted location choice probabilities by individual types. The model's prediction on where people choose to live also aligns well with the data.\footnote{Since we include the residuals in location choice estimation as the exogenous amenity preference in our model, we expect our predicted location choice distributions to match the observed distribution closely. See Figure \ref{fig:Model-Fit-2}.} In subsequent analyses, we will use the estimated model as the benchmark for comparing the effects of various counterfactual experiments.

\section{Trends in Assortative Preference} \label{sec:AM}

% Figure
\begin{figure}[t]
\begin{minipage}[c]{0.99\textwidth}
\includegraphics[height=5.3cm]{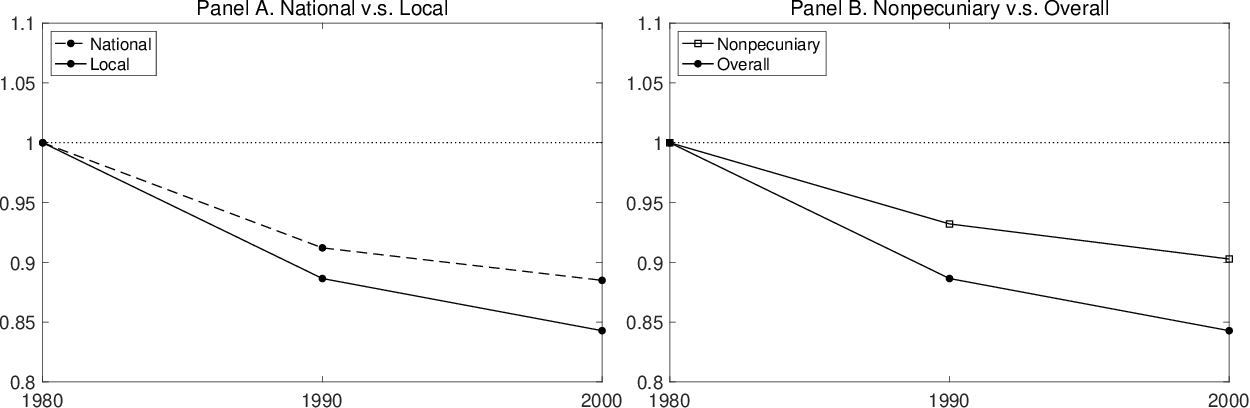}
\caption{Trends of Marital Assortativeness}\label{fig:Trends-of-Marital} 

\vspace{1.0em}
 {\scriptsize{\textit{Notes:} This figure presents the trends of assortative preference measured by the supermodularity of overall marital surplus. Panel A compares the trends based on the national and local marriage assumptions. Panel B compares the nonpecuniary  and the overall assortative preference. The level in 1980 is normalized to be 1. }{\scriptsize\par}
{\scriptsize}}{\scriptsize\par}
\end{minipage}\label{figure:AM_CTAM_Rho}
\end{figure} 
%end

One of the key insights of this paper is that geographic sorting may lead to an increase in the observed homogamy at the national level, even if individual preferences have not changed. Without accounting for educational distributions at the local level, the impact of geographic sorting on marriage could be erroneously attributed to increasing assortative preferences.

In this section, we assess the change in marital assortativeness from 1980 to 2000 by taking local education compositions into account. In addition, our structural model enables us to compute a theoretically grounded measure of marital assortativeness over time. First, let 
\[
\Omega_{mt}\left(e_{M},e_{F}\right)=\sigma^{\epsilon}\left[\overline{V}_{mt}^{e_F}\left(M, e_{M}\right)+\overline{V}_{mt}^{e_M}\left(F, e_{F}\right)-\overline{V}_{mt}^{s}\left(M, e_{M}\right)-\overline{V}_{mt}^{s}\left(F, e_{F}\right)\right]
\]
be the \emph{marital surplus} of the couple with educational levels $\left(e_{M},e_{F}\right)$, which represents the joint utility gains the couple achieves by forming a household over remaining single. Then, following \cite{chiappori2020changes}, we measure assortative preference by the supermodularity core of marital surpluses, defined as
\[
\Gamma_{mt}=\Omega_{mt}\left(H,H\right)+\Omega_{mt}\left(L,L\right)-\Omega_{mt}\left(H,L\right)-\Omega_{mt}\left(L,H\right)
\]
$\Gamma_{mt}$ is the discrete analogue of the cross derivative $\left.\partial\Omega_{mt}\left(e_{M},e_{F}\right)\right/\partial e_{M}\partial e_{F}$ and captures the local degree of assortativeness in marital preferences in marriage market $m$ at time $t$. Since different cities exhibit different marital surpluses due to labor market heterogeneity, $\Gamma_{mt}$ also varies across cities. By averaging across local $\Gamma_{mt}$, we obtain the marital assortativeness at the national level. 

As shown by Figure \ref{fig:Trends-of-Marital}, our analysis reveals a marked decline in overall assortative preferences from 1980 to 2000, by approximately 15.7 percent. This finding contrasts with the benchmark results of \cite{greenwood2014marry} and \cite{eika2019educational}, who report a rising or steady pattern in assortativeness over this period based on reduced-form measures calculated under a national marriage market assumption.\footnote{Several measures in \citet{greenwood2014marry} and \citet{eika2019educational} also show declining assortativeness from 1980 to 2000, notably those based on linear and rank correlation between husbands and wives' education levels.}

To see how the national market assumption matters, Figure \ref{fig:Trends-of-Marital} also presents the trend in assortative preference estimated under the traditional assumption. Marital assortativeness based on national market assumption turns out to have declined by 11.5 percent from 1980 to 2000. Compared to our new estimates  based on local marriage markets, it represents an approximately 27 percent upward bias. This finding suggests that a local market framework is not only theoretically more plausible but also quantitatively meaningful for understanding true shifts in marital preferences.\footnote{In Appendix \ref{ap:am}, we compute reduced-form measures of assortativeness, including Pearson's correlation coefficient and the likelihood ratio index following \cite{greenwood2014marry} and \cite{eika2019educational}. For both measures, we again find that assuming a national marriage market leads to an upward bias in the trend of assortativeness, compared to local marriage market assumptions. In some cases, the conclusions even differ qualitatively. For instance, using Pearson's correlation, the assumption of a national marriage market suggests an increase in assortativeness from 1980 to 2000, while the local market assumption shows a decline. These results highlight that, whether we use reduced-form or structural measures, assuming a national marriage market can significantly skew conclusions about the historical evolution of assortative preferences.}

As will be further analyzed in Section \ref{section:wf}, marital surplus is determined by the pecuniary surplus, the nonpecuniary marital preference, as well as marital transfers between couples. Based on our structural model, we can further explore the driving force of a declining assortativeness. In Panel B of Figure \ref{fig:Trends-of-Marital}, we present results for the evolution in the assortativeness of nonpecuniary preferences. To do so, we compute the supermodularity core of nonpecuniary marital surpluses, defined as
\begin{align*}
\Gamma_{t}^{\mu} & =\left(\mu_{t}^{H}\left(m,H\right)+\mu_{t}^{H}\left(f, H\right)\right)+\left(\mu_{t}^{L}\left(m, L\right)+\mu_{t}^{L}\left(f,L\right)\right)\\
 & \ \ \ \ \left.\left.-\left(\mu_{t}^{L}\left(m,H\right)+\mu_{t}^{H}\left(f,L\right)\right)-\left(\mu_{t}^{H}\left(m,L\right)+\mu_{t}^{L}\left(f,H\right)\right)\right\} \right]
\end{align*}

The results reveal a 9.72 percent decline in the assortativeness of nonpecuniary preferences from 1980 to 2000, suggesting that both financial and non-financial considerations in marriage contributed to the decline in assortative marital preference during this period. Our findings therefore do not support the rising hedonic marriage thesis of \citet{wolfers_comment_2009}, which posits that individuals increasingly prefer spouses with similar educational levels due to a growing emphasis on consumption complementarities. Instead, our results point to a broader shift towards openness in marital preferences, as both pecuniary and nonpecuniary preference assortativeness weakened over time.

\section{Interaction of Local Labor and Marriage Markets} \label{sec:positive}

As documented in Section \ref{sec:desc}, the probability of marrying a high-skill spouse varies significantly across cities and over time.  Based on the estimated model, this section explores how marital preferences and education compositions in each city influence local marital outcomes, as well as how different marital prospects across cities affect individual location choices. 

\subsection{Assortativeness, Spatial Sorting, and College Marital Gap}

% Figure
\begin{figure}[t]
\centering 
\begin{minipage}[c]{0.99\textwidth}
\includegraphics[height=5.9cm]{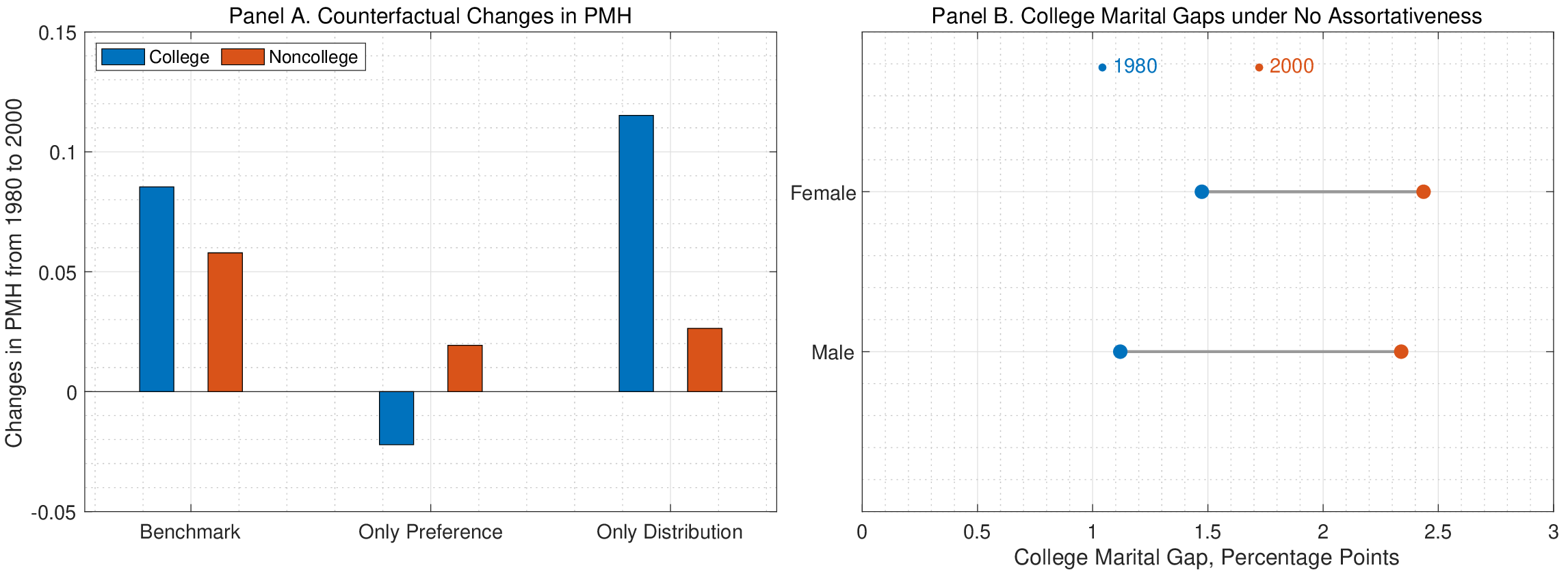}
\caption{College Marital Gap under Counterfactual Experiments}

\vspace{1.0em}
 {\scriptsize\textit{Notes:}}{\scriptsize{} Panel A of this figure presents the changes of the PMH from 1980 to 2000 under the benchmark and the counterfactual experiments. The experiment of `only preference' allows the nonpecuniary marital preference, wages, and rents to change but hold education distributions fixed as in 1980. The experiment of 'only distribution', conversely, only allows education distributions to change. Panel B presents the gap between high-skill and low-skill individuals in the probability of marrying a high-skill partner under random matching. \label{fig:PA_F1} }{\scriptsize\par}%
\end{minipage}
\end{figure}
% end

Nationally, the college marital gap---defined as the difference in the PMH between college and noncollege graduates---widened by 2.75 percentage points from 1980 to 2000, approximately 6\% of the gap in 1980.\footnote{Since this study focuses on assortative marriage, the main marital outcome to look at is the choice for partners with different education levels, conditional on getting married. However, the model accounts for individuals' decisions on whether to get married and shows a good fit.} This widening gap reflects an increase in observed educational homogamy. However, our structural estimates reveal that assortative preferences, both pecuniary and nonpecuniary, have actually declined over this period, suggesting that marital preference is unlikely to be the full story.  Notably, the educational attainment of both men and women rose substantially during this period, coupled with that college graduates increasingly sort into high-skill cities. To disentangle the role of marital preferences and educational distributions in the observed college marital gap, we implement a set of counterfactual experiments using the estimated model. 

In the first experiment, we fix the educational distributions in each city at their 1980 levels, while allowing other factors---including nonpecuniary marital preferences, wages, and rents---to change as they did from 1980 to 2000. These factors shape marital preferences by influencing both the pecuniary and nonpecuniary benefits of marriage.\footnote{While the shifting wages do not fundamentally change individuals' utility functions, i.e. the deep preference parameters, they change an individual's preference for different partners. We slightly abuse the terminology for ease of exposition.} For local marriage markets in each city, we re-solve their equilibria under the counterfactual conditions and compute the resulting college marital gaps. This experiment is partial-equilibrium, holding wages and housing prices as exogenous. Through out following sections, the benchmark represents the model's prediction of the reality.

Panel A of Figure \ref{fig:PA_F1} shows that, comparing the benchmark with the counterfactual experiment that only allows preference change, the decline in assortative preferences would lead to a narrowing of the national college marital gap by 4.14 percentage points. Conversely, if we fix wages, rents, and nonpecuniary preferences at their 1980 levels and only allow the educational distributions to evolve as they did, the national college marital gap would have widen by 8.88 percentage points. These complementary findings suggest that the observed increase in national marital gap was primarily driven by the shifts in educational compositions over time, rather than changes in marital preferences.

However, despite the assortative preference has declined over time, it plays nontrivial roles in widening the college marital gap by interacting with the shifting education distributions. Note that the educational composition of cities evolved over time due to two factors: the overall rise in educational attainment and the geographic sorting of high-skill individuals into high-skill cities. The first factor---the increase in educational attainment particularly among women---has led to a significant expansion of the college-educated population. If individuals exhibited no assortative preference, this increasing prevalence of college graduates would lead to an equal increase in the PMH between college and non-college graduates. However, under assortative preference, it disproportionately raise the likelihood of college-graduates marrying one another, resulting in a widening national college marital gap.

Furthermore, geographic sorting exacerbates the college marital gap as the second factor. As high-skill individuals increasingly cluster in high-skill cities, their local marriage pools become more concentrated with similarly educated potential partners, increasing their probability of marrying within their educational group, even if they exhibited no assortative preferences. To single out the role of spacial sorting, we conduct a counterfactual experiment assuming random matching, which effectively sets assortative preferences to ``zero,'' while allowing the educational compositions of cities to evolve as observed from 1980 to 2000.
 
Panel B of Figure \ref{fig:PA_F1} presents the results. Although exhibiting no assortative preference, high-skill individuals turn out to have higher PMH, with the national college marital gap standing at 1.1 percentage points for male and 1.6 percentage points for female in 1980. This gap arises from pre-existing spatial sorting: high-skill individuals tend to live in high-skill cities, where the local marriage pool is richer in college-educated partners. Over time, as geographic sorting intensified, the national gap under random matching widened to 2.4 and 2.6 percentage points for men and women in 2000. This experiment suggests that geographic sorting has played a nontrivial role in increasing observed homogamy at the national level. In particular, the impacts of geographic sorting on marriage matching can be mistakenly attributed to a rising assortative preference if local marriage markets are not accounted for. 

Our analysis so far focuses on college marital gap at the national level, but marital outcomes exhibit substantial variation across cities. We also examine how changes in local preferences  and educational distributions have shaped the spatial inequality in marital outcomes, measured by the Gini coefficient in local PMH. The results reveal a nuanced picture: Over time, the evolution of assortative preferences increased the spatial inequality in marital outcomes for college graduates while reducing it for noncollege graduates. Conversely, changes in the educational compositions of cities---particularly the rising share of college-educated individuals---widened the spatial inequality in marital outcomes for noncollege graduates but narrowed it for college graduates.

\subsection{Marital Prospects and Location Choices}

% Figure
\begin{figure}[t]
\centering 
\begin{minipage}[c]{0.99\textwidth}
\includegraphics[height=5.9cm]{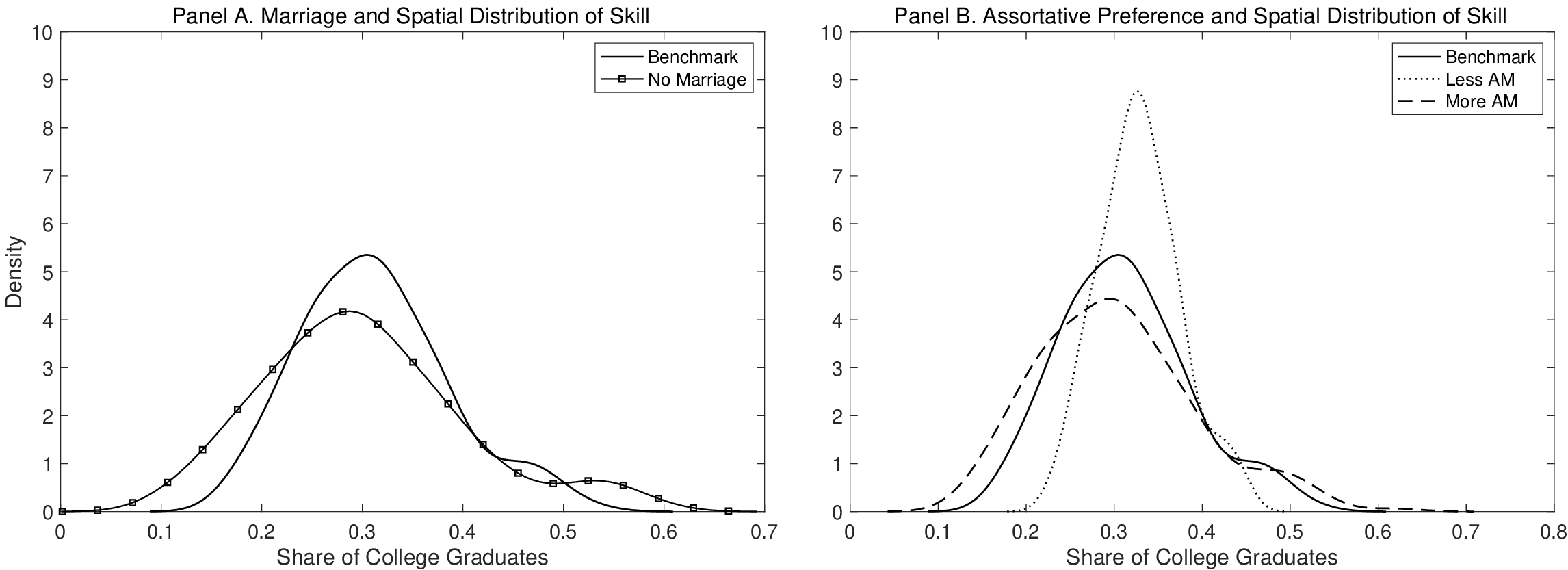} \caption{Spatial Distribution of Skill under Counterfactual Experiments}

\vspace{1em}
 {\scriptsize\textit{Notes:}}{\scriptsize{} Panel A of this figure shows the distribution of the share of college graduates across cities, with and without marriage. Panel B presents the distributions under different strengths of assortative preference. \label{fig:PA_F2}}{\scriptsize\par}%
\end{minipage}
\end{figure}
% end

While we have shown that where individuals live influences with whom they marry, with whom individuals  could marry may also influence where they choose to live. The significant geographic heterogeneity in local population, wage, and marital outcomes implies that potential marital gains can differ substantially across cities. In what follows, we consider how marital prospects across cities shape individuals' location choices. The counterfactual experiments find that marriage \textit{per se} makes high-skill cities less skill intensive. However, we further reveal that assortative preference has a reinforcing effect on spatial sorting.

Our first exercise simulates individuals' location choices in 2000 with a hypothetical scenario in which marriage incurs infinite negative utility. Under this exercise, everyone remains single and marriage becomes an infeasible option. Panel A of Figure \ref{fig:PA_F2} presents the result, showing that the absence of marriage option would lead to greater geographical dispersion in skills. The share of college graduates in the upper quartile city was 9.55 percentage points higher than the lower quartile city, but it is widened  to 12.41 percentage points when marriage becomes infeasible, an increase of 29.9\%. This finding aligns with \cite{Fan2021The}, who shows that marriage \textit{per se} facilitates a more even spatial distribution of skills across cities. As will be analyzed in Section \ref{section:wf}, the underlying reason is that, compared to noncollege graduates, marital surplus for college graduates in high-skill cities is disproportionately smaller  than in low-skill cities.

As high-skill individuals prefer marrying alike, would they sort into high-skill cities for marital prospects? Next, we examine the role of assortative preference by simulating counterfactual location choices under different degree of assortativeness. According to our structural model, the extent to which individuals  marry assortatively hings on both the pecuniary and nonpecuniary marital preferences.  Since the pecuniary preference furthers depend on wage structures across genders and education levels, we manipulate assortative preference by changing the value of the nonpecuniary utility parameter $\mu$, as it is exogenous primitive of the structural model. Specifically, we increase the nonpecuniary marital surpluses of couples with the same education levels ( $\mu_{HH}$ and $\mu_{LL}$), while reduce the surpluses of couples with different education levels ($\mu_{HL}$ and $\mu_{LH}$), so that the nonpecuniary supermodularity core, defined as $\mu_{HH}+\mu_{LL}-\mu_{HL}-\mu_{LH}$, becomes stronger.

The results, presented in Figure \ref{fig:PA_F2}(b), show that a stronger preference for assortative marriage would lead to notable increase in spatial inequality of skills. With 50 percent stronger nonpecuniary assortative preference, the gap in the share of college graduates between the upper and lower quartile would rise from 9.55 to 11.33 percentage points by 18.6 percent. Conversely, when individuals exhibit weaker assortative preferences, college graduates would distribute more evenly across cities. Although it is intuitive that, with stronger preference for assortative marriage, high-skill individuals would want to live in high-skill cities as there are more marital candidates with desired type. However, individuals would also face stronger competitions from peers of the same gender. Therefore, it is important to capture this equilibrium effect in local marry markets. According to our structural model, a larger supply of college graduates of the opposite gender in a high-skill city would lead to larger marital transfers from them. Meanwhile, the transfer would be reduced by a larger supply of college graduates of own gender. Our experiments capture this equilibrium effect by re-solving the equilibrium of each local marriage market under new counterfactual assortative preferences.

Finally, taking the trends into account, the above findings---marriage \textit{pe se} and assortative marital preference having opposite impacts on location choices---also suggest opposite impacts on the evolution of spatial sorting over time. The lowering marriage likelihood from 1980 to 2000 has contributed to the rising spatial sorting, whereas a declining assortative preference has concurrently mitigated this impact.

\section{Implications on Household Inequality}

% Table
\begin{table}[t]
\centering
\footnotesize 
\caption{Household Inequality}
\label{tab:ineq01}
\begin{minipage}{0.99\textwidth} 
\tabcolsep=6pt 
\begin{tabular}{p{0.25\textwidth}*{3}{>{\centering}p{0.21\textwidth}}} 
\toprule
\toprule
Year & National & Local, Mean & Local, Std. Dev. \tabularnewline
\midrule
1980 & 0.120 & 0.094 & 0.019 \tabularnewline
2000 & 0.164 & 0.139 & 0.022 \tabularnewline
\midrule
Pct. Change & 37.25\% & 47.13\% & 16.11\% \tabularnewline
\bottomrule
\end{tabular}

\vspace{0.5em} 
\scriptsize 
\noindent\textit{Notes:}
Household inequality is measured by the Gini coefficient with household equivalence scale $\left(1+\chi\right)=1.7$. Column (1) reports the national household inequality level. Column (2) and (3) report respectively the mean and standard deviation of local household inequality.
\end{minipage}
\end{table}
% end

From 1980 to 2000, household inequality rose significantly in the U.S, as measured by the Gini coefficient. In the meantime, the dispersion in local inequality also increased. Some cities became significantly more unequal relative to others. The U.S. was becoming more unequal both nationally and in its local inequality levels during this period of time, as shown in Table \ref{tab:ineq01}.

Traditional literature has emphasized the role of changing wage structure: the rising college premium, driven by skill-biased technological change, raised inequality, whereas the closing gender wage gap, driven by decreasing discrimination in the workplace, helped shrink it \citep{autor_trends_2008, lee_accounting_2010}. However, changes in wage inequality do not fully explain changes in household inequality, as family formation has been largely ignored or taken as exogenous. 

In this section, we examine how assortative marriage, together with geographic sorting, affects household inequality and quantify how much these interlinked phenomena contributed to the observed rise in inequality at both the national and the local levels.

\subsection{Assortative Marriage and National Inequality}

% Table
\begin{table}[t]
\centering
\footnotesize 
\caption{Household Inequality Under Counterfactual Experiments}
\label{tab:ineq02}
\begin{minipage}{0.99\textwidth} 
\tabcolsep=6pt %
\begin{tabular}{p{0.15\textwidth}*{6}{>{\centering}p{0.11\textwidth}}} 
\toprule
\toprule
 & \multicolumn{2}{c}{(1)} & \multicolumn{2}{c}{(2)} & (3) & (4) \tabularnewline
\cmidrule(lr){2-3} \cmidrule(lr){4-5}
 & RM & AM & RM & AM &  & \tabularnewline
\midrule
1980 & 0.113 & 0.120 & 0.113 & 0.120 & 0.120 & 0.120 \tabularnewline
2000 & 0.114 & 0.123 & 0.144 & 0.152 & 0.121 & 0.122 \tabularnewline
\midrule
Pct. Change & 0.41\% & 2.77\% & 27.17\% & 27.47\% & 1.51\% & 1.63\% \tabularnewline
\bottomrule
\end{tabular}

\vspace{0.5em} 
\scriptsize 
\noindent\textit{Notes:} RM refers to random matching, where households are formed randomly. AM refers to assortative marriage, where households are formed based on observed assortative preferences. In Experiment (1), the educational level in each city is adjusted to reflect the national increase in educational attainment from 1980 to 2000. Experiment (2) adjusts the college premium in each city to reflect the national rise during the same period. Experiment (3) allows individuals' location choice probabilities to evolve from their 1980 levels to their 2000 levels. Finally, Experiment (4) incorporates both changes in location choice probabilities and city wage differentials from 1980 to 2000. In all experiments, factors not explicitly mentioned are held constant at their 1980 levels.
\end{minipage}
\end{table}
% end

We first conduct a simple exercise based on \cite{greenwood2014marry} and \cite{eika2019educational}, in which we compute the household inequality that would have been realized if individuals had married randomly rather than assortatively, taking their locations, income, and probability of getting married as given. In contrast to \cite{greenwood2014marry} and \cite{eika2019educational}, our exercise is based on local marriage markets and accounts for the differences in local educational compositions.

We find that in 1980, if people married randomly in each city, the national inequality would have been 4.9 percent lower. Moreover, random marriage would result in a 7.4 percent decline in national inequality in 2000. Therefore, assortative marriage not only amplifies household inequality, but has also contributed to its increasing trend over time. A simple calculation shows this contribution to be 4.2 percent between 1980 and 2000.

Why would assortative marriage play a greater role in household inequality, despite that its trend has actually declined? First, when individuals marry assortatively, more households are formed with both college-educated partners or with both non-college-educated partners. Over time, as the college premium rises, the income gap between these high-skill and low-skill families would increase. Second, as revealed in the previous section, rising educational attainment over time interacts with assortative preferences to widen the college marital gap and generate greater educational homogamy. This, in turn, would further exacerbate household income disparities and intensify overall inequality.

To demonstrate these two channels, Panel (1) and (2) of Table \ref{tab:ineq02} present the results of our second set of exercises, in which we fix all factors at their 1980 levels, while changing only the \textit{national} educational level or the college premium. In experiment (1), where only the educational level increases, household inequality under assortative marriage rises by 2.8 percent, compared to a much smaller increase of 0.4 percent under random matching. This result highlights the role of assortative marriage in amplifying inequality when educational attainment rises. In experiment (2), where only the college premium rises, inequality grows significantly, increasing by 27.5 percent under assortative marriage and 27.2 percent under random matching. These results indicate that rising college premium drives household inequality primarily through its direct effect, with marital sorting playing a smaller, albeit still meaningful, role.

These findings underscore the non-ignorable role of assortative marriage to rising household inequality. Notably, although marital assortativeness slightly declined from 1980 to 2000---a trend that could have mitigated inequality growth---our results demonstrate that the persistent level of assortative preferences has had a more substantial impact on increasing inequality over time. This effect arises through its interactions with shifting educational distributions and wage structures. These insights align with the findings of \citet{eika2019educational}, who reveal that it is the level of marital assortativeness, rather than its change over time, that has driven its impact on household inequality. Our study further highlights its influence through the interaction with rising educational attainment and skill premiums.

\subsection{Spatial Sorting and National Inequality}

Spatially, however, wage and education levels grow unevenly. The process of geographic sorting and the phenomenon of the great divergence---whereby high-skill cities become increasingly high-wage and high-skill---can further raise household inequality at the national level. First, as demonstrated in Section \ref{sec:positive}, geographic sorting acts as a catalyst for educational homogamy. As individuals with similar educational levels increasingly reside in the same areas, the likelihood of marrying someone of similar educational attainment increases even under random matching. Consequently, geographic sorting reinforces educational homogamy and exacerbate household inequality at the national level. Second, high-skill cities also experienced faster wage growth during this period. While this process by itself would increase national inequality, it is exacerbated by assortative marriage, which generates more households with both high-skill spouses concentrated in these cities.

To quantify the impacts of spatial sorting and the great divergence under assortative marriage, we implement our third set of experiments. In Table \ref{tab:ineq02} Column (3), we allow individuals' location choice probabilities to evolve over time, while fixing everything else at 1980 levels. This exercise looks at how national household inequality would change if high-skill and low-skill individuals in 1980 were sorted across cities as they did in 2000. The result suggests that geographic sorting alone would lead to a 1.5 percent increase in national inequality, which represents 4 percent of the national inequality growth from 1980 to 2000. Table \ref{tab:ineq02} column (4) implements a counterfactual experiment in which we allow both individuals' location choice probabilities and the wage differentials among cities to evolve over time, holding everything else (including the national wage level) at their 1980 levels. The exercise reveals the impact of the great divergence: under assortative marriage, the uneven spatial growth in skills and wage levels together raised national inequality by 1.6 percent, accounting for 4.5 percent of national inequality growth from 1980 to 2000.

\subsection{Local Inequality}

% Figure
\begin{figure}[t]
\centering
\begin{minipage}[c]{0.99\textwidth} 
\centering
\includegraphics[width=0.99\textwidth]{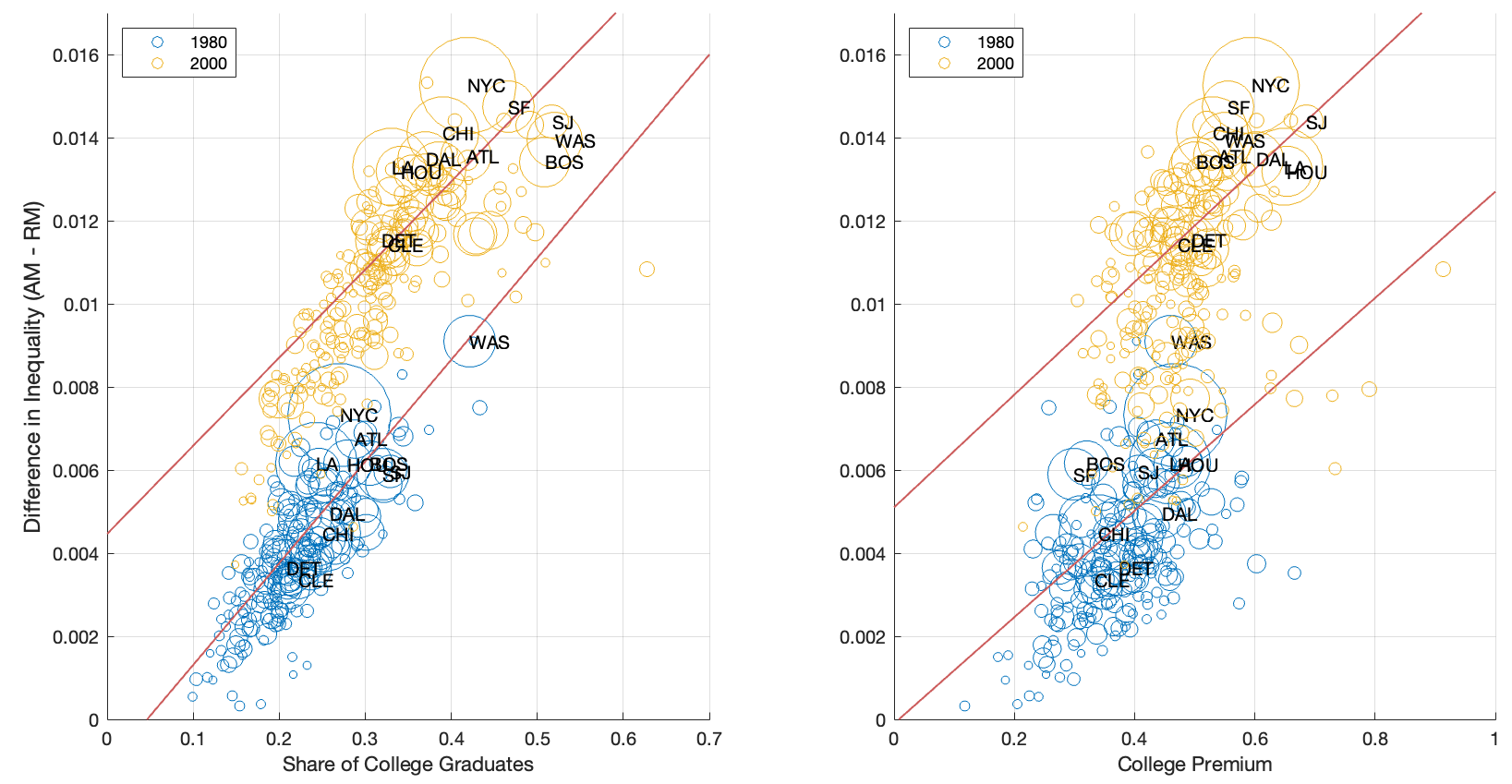} 
\caption{Education Level, College Premium, and Household Inequality}
\label{fig:ineq01}

\vspace{1.0em} 
\raggedright
\scriptsize 
\noindent\textit{Notes:} This plot shows the relation between the share of college graduates (left panel), the college premium (right panel), and the difference in local household inequality under assortative marriage (AM) and random matching (RM).
\end{minipage}
\end{figure}
% end

While assortative marriage contributed to rising national household inequality over time, the contribution was not uniform across space. Figure \ref{fig:ineq01} highlights the heterogeneous role of assortative marriage across cities by plotting the difference in local household inequality under assortative marriage and random marriage against respectively a city's college graduate share and college premium. The figure reveals that cities with more college graduates and higher college premiums experienced larger increases in inequality levels due to marital assortativeness---consistent with our previous finding on how marital assortativeness interacts with shifting educational distributions and skill premiums to drive inequality at the national level.

Local inequalities matter since they have consequences that extend beyond income disparities. It affects various aspects of communities. Cities with high household inequality often feature lower intergenerational mobility, higher levels of neighborhood segregation and crime rates, unequal access to amenities, and uneven distribution of education and health care resources \citep{glaeser2009inequality}. Taken together, the left and right panels of Figure \ref{fig:ineq01} paint a picture of how assortative marriage contributed to increasing inequality at the local level, particularly for cities with high educational level and wage inequality. In 1980, if people married randomly in NYC, for example, inequality in the city would have declined by 5.4 percent. By 2000, random marriage would have reduced inequality by 8.4 percent, even while marriage rates declined. In contrast, for residents in Johnstown PA, random marriage would cause inequality to decline by a much smaller 2.8 percent and 4.4 percent in 1980 and 2000. Overall, marital assortativeness accounted for 12 percent and 15 percent of the variations in local inequality levels across the U.S. in these two periods.

In conclusion, the intertwining relationship between assortative marriage and geographic sorting has been a driving force behind the increasing household inequality in the U.S. Assortative marriage concentrates income within specific households, while geographic sorting further amplifies assortative marital patterns by clustering individuals with similar educational backgrounds in particular regions. Overtime, as high-skill cities became increasingly high-wage and high-skill, this interplay between assortative marriage and geographic sorting exacerbates household inequality both nationally and locally.

\section{Welfare Analysis}
\label{section:wf}

The impacts of geographic sorting and assortative marriage on inequality are primarily analyzed through the lens of pecuniary outcomes. A key feature of marriage, however, is that it matters for people's well-being beyond just financial aspects. What, then, are the general welfare implications of geographic sorting and assortative mating? This section seeks to explore answers to this question.

\subsection{Location and Marital Surplus}

% Figure
\begin{figure}[t]
\centering 
\begin{minipage}[c]{0.99\textwidth}
\includegraphics[height=10cm]{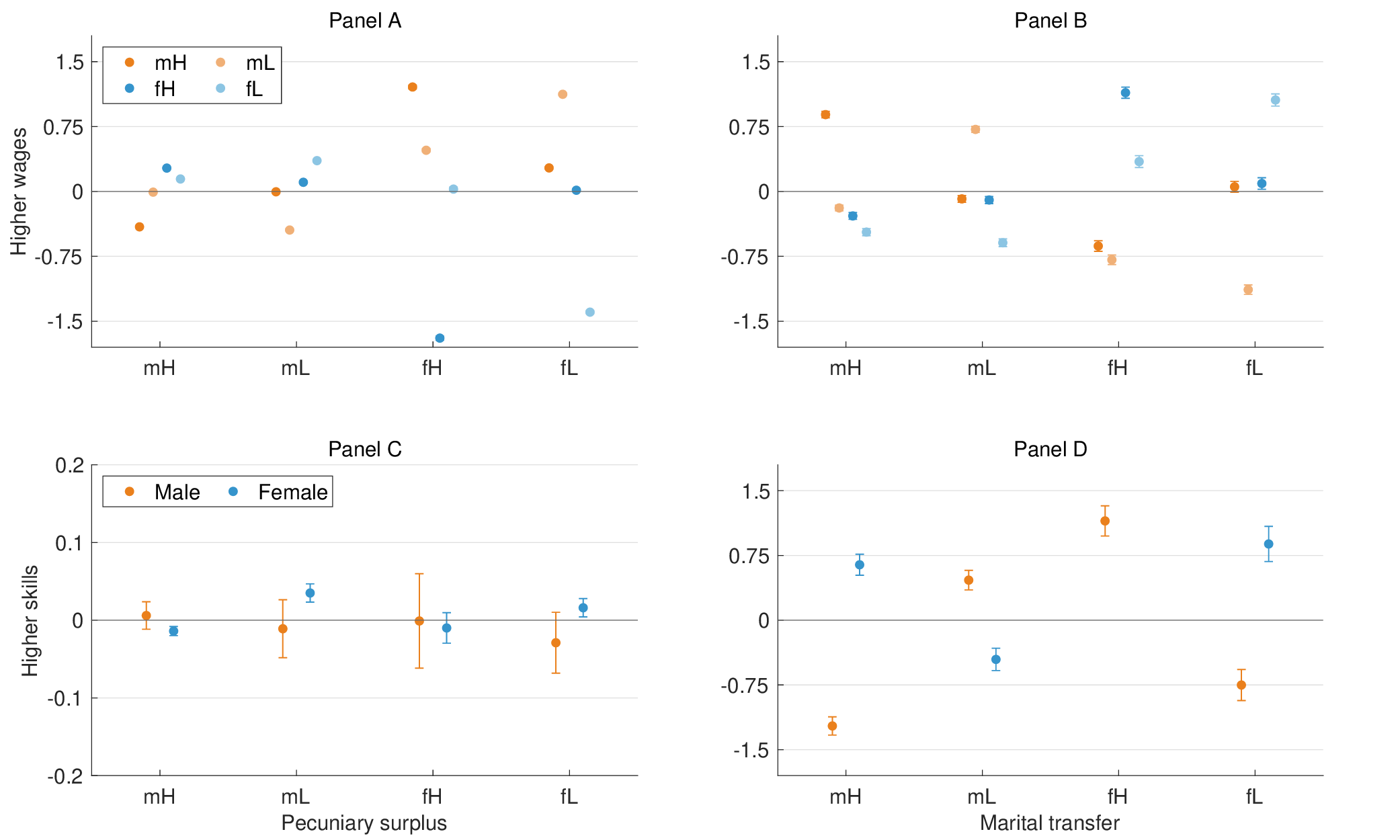} \caption{Effects of Wages and Education Distributions on Local Marital Surplus}

\vspace{1em}
 {\scriptsize\textit{Notes:}}{\scriptsize{} This figure presents the estimates from regressions of the pecuniary marital surplus and marital transfers on local shares of college graduates and local log wages.  By four different types of individuals, panels on the left (A and C) present the effects on their pecuniary surplus, while panels on the right (B and D) present the effects on their marital transfers. Panels on the top (A and B) show the effects of higher log wages, while panels  on the bottom (C and D) show the effects of larger share of college graduates. Time fixed effects are controlled. mH: high-skill male; mL: low-skill male; fH: high-skill female; fL: low-skill female. 95\% confidence intervals are presented. \label{fig:WF_F1}}{\scriptsize\par}%

\label{fig:wf1}
 
\end{minipage}
\end{figure}
% end

 Given the significant geographic heterogeneity not only in local labor markets but also in local marriage markets, how does individuals' well-being differ across cities by taking marriage into account? Living in high-skill cities enjoys a higher labor income, which implies that the pecuniary benefits from marriage due to economy of scale may be smaller as compared to remaining single. However, the higher wages in high-skill cities may also have additional effects through local marriage market equilibrium: despite that higher wages imply lower pecuniary marital benefits, the individual will also become more favorable in local marriage market and receive more marital transfers. In addition, these effects are further complicated by the fact that the labor income of potential partner also tends to be higher in high-skill cities.

 To diagnose the welfare implications of marriage and location, we focus on how marital surpluses---the additional utility gain under marriage as compared to being single---vary with local wages and skills across cities. According to our structural model, marital surplus can be decomposed into three components: the pecuniary surplus, the nonpecuniary surplus, and the marital transfer. Our framework based on local  markets captures the rich variation in these components both across cities and over time, allowing us to empirically estimate their relationships and conduct statistical inference.

Based on our structural model, we first recover the overall marital surplus as well as its each component, which are specific to each city. Details for this implementation has been outlined in Section \ref{section:ei}. To provide a quantitative summary of the relationship between marital surpluses, wages and education distributions, we then estimate regressions of each component of marital surplus on the share of college graduates and the wage levels of different individual types, exploiting variation across cities and over time. Welfare measures are converted into monetary units in the natural log dollar points.\footnote{Based on the linear-log indirect utility functions, both the pecuniary surplus and the transfer are mapped to log wages linearly with coefficient 1.}

 Figure \ref{fig:wf1} presents the regression results, which are generally consistent with the intuition: cities with higher income of a given individual type correspond with lower pecuniary marital surplus for these individuals (Panel A), but local marriage market equilibrium leads to nontrivial counter effect by increasing the marital transfer they receive (Panel B). More importantly, individuals of different educational levels from the same city - while earning different incomes from their work - further obtain distinct gains from their marriage due to assortative marriage. For instance, as revealed in Panel A, living in cities with higher wages of high-skill males, high-skill females benefit substantially more in the pecuniary marital surplus compared to those low-skill females. In contrast, living in cities with higher wages of low-skill males, the benefits are notably opposite. 
 
Local education composition also generates uneven welfare impacts by interacting with assortative marital preference.  As revealed by Panel D, high-skill males would benefit from marriage by receiving more marital transfers when living in a city with more high-skill females. However, low-skill males would actually receive less transfers when living in this city, compared to other cities with fewer high-skill females.

Because college wage premiums, gender wage gaps, as well as education compositions differ across cities, how does individual's well-being vary across cities is an empirical question. We further regress the marital surplus combining each component on local share of college graduates and wage levels to have an overall assessment, finding that high-skill individuals benefit less from marriage by living in high-skill cities compared to low-skill individuals. This is consistent with our previous findings that marriage reduces spatial dispersion of skills.

\subsection{The Evolution of the College Welfare Gap}

% Table
\begin{table}[t]
  \footnotesize
  \centering
  \begin{minipage}{0.99\textwidth}
   \caption{The Evoluation of College Welfare Gap from 1980 to 2000}
   \tabcolsep=4pt
   \begin{tabular}{p{0.28\textwidth}>{\centering}p{0.1\textwidth}>{\centering}p{0.1\textwidth}>{\centering}p{0.1\textwidth}>{\centering}p{0.1\textwidth}>{\centering}p{0.1\textwidth}>{\centering\arraybackslash}p{0.1\textwidth}}
    \hline\hline
    & (1) & (2) & (3) & (4) & (5) & (6) \\
    \hline
    1980       & -0.051 & -0.051 & -0.051 & -0.051 &  -0.051    &  -0.051       \\
    2000       & 0.111  & 0.097  & 0.117  & 0.234  &   0.239    &   0.204    \\
    1980--2000  & 0.161  & 0.147  & 0.168  & 0.285  &   0.289    &   0.255    \\
    \hline
    Wage       & Yes    & Yes    & Yes    &   Yes     &   Yes    &   Yes    \\
    Rent       &     &     Yes   & Yes    &   Yes     &   Yes    &   Yes    \\
    Nonpecuniary Utility  &        &        &        &    Yes    &    Yes   &   Yes    \\
     Location Choice &        &        &        &        &  Yes    &   Yes    \\
           Education Composition &        &        &        &        &      &   Yes    \\         
    \hline
        Marriage Market Equilibrium       &   No     &    No    &   Yes     &   Yes     &   Yes    &   Yes    \\
      \hline  
   \end{tabular}

   \vspace{1.0em}     
   \scriptsize{ \textit{Notes:} This table presents changes in the college welfare gap from 1980 to 2000, based on counterfactual experiments sequentially allowing more factors to change over time. Column (1) allows only wages to change, while all other factors, including marital transfers, remain fixed at their 1980 levels. Column (2) adds changing rents to wages, without re-solving marriage market equilibria. Column (3) re-solves marriage market equilibria, allowing wages and rents to affect marital transfers. Column (4) further includes changes in nonpecuniary utilities. Column (5) incorporates changing location choice probabilities, and Column (6) allows the marginal distributions of education levels to change as well. From column (3) onward, marital transfers adjust as part of the equilibria, rather than remaining fixed.}   
   \label{table:wg_all}
  \end{minipage}
 \end{table}
 % end

Over time, as individuals increasingly sorted across locations by skill, and local labor and marriage market conditions evolved, welfare inequality between high-skill and low-skill individuals also changed. Urban studies  highlight the critical role of location in shaping the welfare gap between college and noncollege graduates. While college premium---the wage gap between college and noncollege graduates---has steadily increased, \cite{moretti2013real} argues that the welfare gap has grown more slowly when real wages are considered. This is because college graduates tend to sort into high-skill, high-wage cities, but higher living costs offset their nominal wage gains. In contrast, \cite{diamond2016determinants} demonstrates that the college welfare gap has outpaced the nominal wage gap when urban amenities are taken into account. As more high-skill workers cluster in a city, urban amenities improve, further enhancing the relative welfare of college graduates.

In this section, we add to these studies, which are only based on the sample of male workers, by taking household as well as its formation into account.\footnote{Both \cite{moretti2013real} and \cite{diamond2016determinants} study the sample of male workers only.} Marriage introduces additional complexities: as college graduates increasingly sort into high-skill cities and form households with similarly educated spouses, the pecuniary surplus, nonpecuniary surplus, and marital transfers they receive, relative to noncollege graduates, will determine whether the college welfare gap widens further beyond the real wage gap. To assess the welfare implications of this interaction between geographic sorting and assortative marriage, we simulate college welfare gap by sequentially altering underlying factors, in line with the analysis by \cite{diamond2016determinants}.

Table \ref{table:wg_all} presents the evolution of the national college welfare gap under counterfactual scenarios.  Column (1) only allows local wages to change from 1980 to 2000, keeping all other factors fixed at their 1980 levels. Column (2) adds changing rents in addition to wages. Being consistent with previous studies, the result shows that the widening college welfare gap becomes smaller once housing costs are considered, dropping from 16.1 to 14.7 log points, highlighting the offsetting role of higher living costs in high-wage cities.\footnote{Our results based on single males, presented by Appendix Table \ref{table:wg_m}, are also both qualitatively and quantitatively comparable with the results by \cite{moretti2013real} and \cite{diamond2016determinants}.}

However, these effects are quantified without considering marriage market equilibrium, where marital transfers between couples are held fixed. In practice, as we have shown in the previous subsection, increasing skill premiums also lead to increased equilibrium marital transfers. In column (3), we allow marital transfers to change endogenously in response to shifting wages and rents. As a result, the college welfare gap further widens by about 2 log points, reflecting how high-skill individuals become more desirable in marriage markets and receive greater transfers in equilibrium. Further decomposition by gender in Appendix Tables \ref{table:wg_m} and \ref{table:wg_f} reveals that this widening welfare gap is primarily driven by males. Both high- and low-skill males receive fewer transfers due to the narrowing gender wage gap, but low-skill males are disproportionately affected as the rising skill premium reduces their relative desirability in marriage markets.

Beyond pecuniary benefits, nonpecuniary marital preferences also play a role in shaping the college welfare gap. Column (4) incorporates the changes in nonpecuniary utilities, in addition to wages, rents and re-solving for marriage markets equilibria. The college welfare gap substantially increases to 28.5 log points, reflecting that high-skill individuals gain more from marriage beyond economic factors, consistent with our structural estimates of $\mu$.\footnote{Note that the nonpecuniary surplus is relative to being single. Therefore, an alternative explanation is that low-skill individuals benefit increasingly less from marriage over time.}

To directly gauge the role of spatial sorting, Column (5) further allows location choice probabilities to evolve over time, increasing the college welfare gap by 0.4 log points. While modest, this effect represents the additional contribution of sorting after accounting for local wages, rents, and marital surplus changes. 

Finally, Column (6) allows the marginal distributions of education to change as well, including the rising overall education attainment over this period of time. The result shows that, after considering the increased supply of high-skill individuals, the widening college welfare gap moderates but remains significantly larger than when considering wages and rents alone. As the share of high-skill individuals increases, they become more common in marriage markets, reducing the transfers they receive. Further decomposition by gender in Appendix Tables \ref{table:wg_m} and \ref{table:wg_f} reveals that this effect is primarily driven by females due to the significant improvement in their education. For males, rising educational levels actually widens the college welfare gap, mainly due to assortative marriage. As the increase in high-skill females outpaces that of males, assortative marriage ensures that high-skill males disproportionately benefit from the growing pool of high-skill spouses.

In conclusion, the college welfare gap widened substantially over the two decades from 1980 to 2000, driven not only by rising skill premiums but also by the interplay of nonpecuniary marital preferences, geographic sorting, and changes in educational attainment, particularly once assortative family formation is accounted for. Our analysis complements that of \cite{diamond2016determinants} by emphasizing the forces beyond real wages that shape the evolving welfare gap between college and noncollege graduates.

\section{Conclusion}

This paper investigates the interconnected roles of marital and geographic sorting in shaping marriage patterns, household inequality, and welfare in the United States from 1980 to 2000. We develop and estimate a spatial equilibrium model that nests a marriage matching model under transferable utility, incorporating local marriage markets in addition to local labor and housing markets. Our analysis reveals several key insights. First, while marital preferences remained highly assortative, they did not intensify during this period. Nonetheless, rising educational attainment and geographic sorting interacted with assortative preferences to significantly increase homogamy and household inequality at the national level. Additionally, we find that assuming a national marriage market leads to an overestimation of trends in assortative preferences by attributing increases in homogamy, driven by geographic sorting, to rising assortativeness. This underscores the importance of incorporating local marriage markets into the analysis.

We also analyze the welfare implications of marital and geographic sorting. As
college graduates increasingly resided in high-skill, high-wage cities and married similarly
educated spouses, we find that the welfare gap between college and non-college workers
widened substantially more than their real wage gap. These findings underscore the necessity of jointly modeling labor and marriage markets to comprehensively understand the
socioeconomic dynamics of marriage, inequality, and welfare.
% References 
\newpage
\bibliographystyle{apalike}
\bibliography{Draft_v03}

% Appendix
\newpage
\appendix

\titleformat{\section}
  {\normalfont\Large\bfseries}
  {Appendix \thesection.}{0.5em}{}

\section{Additional Estimates and Model Fits}
\label{ap:fit}

This section provides additional results of the estimation and model fits. Table \ref{tab:est_m} shows the estimates for $\mu$ of each type of individuals by marrying spouse with different education levels.

% Table
\begin{table}[H] 
\centering
\footnotesize 
\caption{Parameter Estimates: Nonpecuniary Utility for Marriage Partners}
\label{tab:est_m}

\begin{minipage}{0.99\textwidth} 
\centering
\tabcolsep=6pt 

\begin{tabular}{p{0.25\textwidth}*{6}{>{\centering}p{0.10\textwidth}}} 
\toprule 
\toprule
 & \multicolumn{3}{c}{High-skill} & \multicolumn{3}{c}{Low-skill} \tabularnewline
\cmidrule(lr){2-4} \cmidrule(lr){5-7}
Self\textbackslash Spouse & 1980 & 1990 & 2000 & 1980 & 1990 & 2000 \tabularnewline
\midrule 
High-skill Male   & 0.558 & 0.478 & 0.445 & 0.663 & 0.556 & 0.525 \tabularnewline
                  & (0.313) & (0.307) & (0.306) & (0.320) & (0.317) & (0.316) \tabularnewline
Low-skill Male    & 0.101 & -0.100 & -0.088 & 0.832 & 0.560 & 0.555 \tabularnewline
                  & (0.309) & (0.303) & (0.303) & (0.317) & (0.311) & (0.310) \tabularnewline
High-skill Female & -0.110 & -0.075 & 0.026 & -0.132 & 0.103 & 0.196 \tabularnewline
                  & (0.312) & (0.308) & (0.306) & (0.304) & (0.304) & (0.305) \tabularnewline
Low-skill Female  & -0.400 & -0.474 & -0.534 & 0.204 & 0.287 & 0.201 \tabularnewline
                  & (0.331) & (0.330) & (0.330) & (0.312) & (0.307) & (0.306) \tabularnewline
\bottomrule 
\end{tabular}

\vspace{0.5em} 

\raggedright
\scriptsize 
\noindent\textit{Notes:} Estimates are relative to the nonpecuniary utilities of being single, which is normalized as zero. Standard errors are in parenthesis.

\end{minipage}
\end{table}
% end

\begin{figure}[H]
\centering 
\begin{minipage}[c]{0.99\textwidth} 
\centering
\includegraphics[width=0.95\columnwidth]{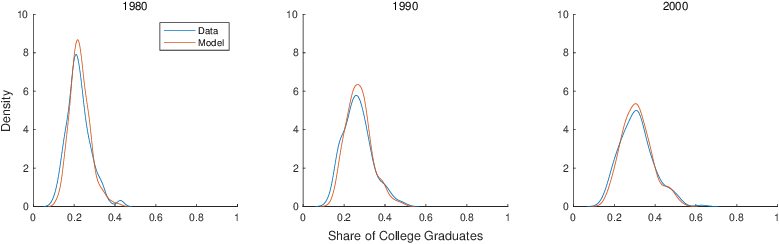} 
\caption{Model Fit---Spatial Distributions of Data and Predicted Share of College Graduates}
\label{fig:Model-Fit-L-1}

\vspace{1.0em} 
\raggedright
\scriptsize 
\noindent\parbox{1.0\textwidth}{
    \textit{Notes:} This figure compares the observed share of college graduates with the model's prediction by their spatial distributions across cities.
    }
\end{minipage}
\end{figure}

\begin{figure}[H]
\centering 
\begin{minipage}[c]{0.99\textwidth} 
\centering
\includegraphics[width=0.95\columnwidth]{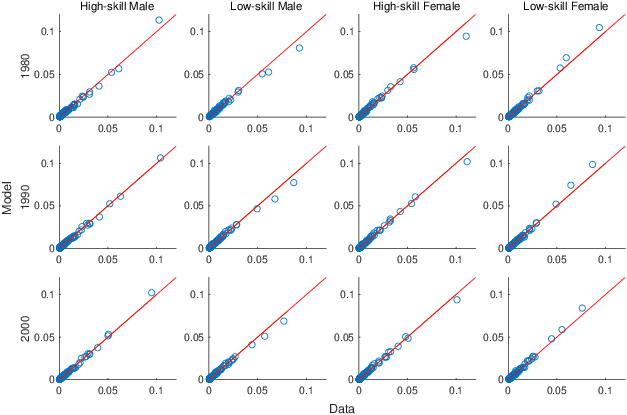} 
\caption{Model Fit---Scatter Plots of Data and Predicted Location Choices}
\label{fig:Model-Fit-L-2}

\vspace{1.0em} 
\raggedright
\scriptsize 
\noindent\parbox{1.0\textwidth}{
    \textit{Notes:} This figure compares the observed location choice probabilities with the model's prediction by scatter plot, with a 45-degree reference line. Each data point represents the choice probabilities for a city.
    }
    
\end{minipage}
\end{figure}

\begin{figure}[H]
\centering 
\begin{minipage}[c]{0.99\textwidth} 
\centering
\includegraphics[width=0.95\columnwidth]{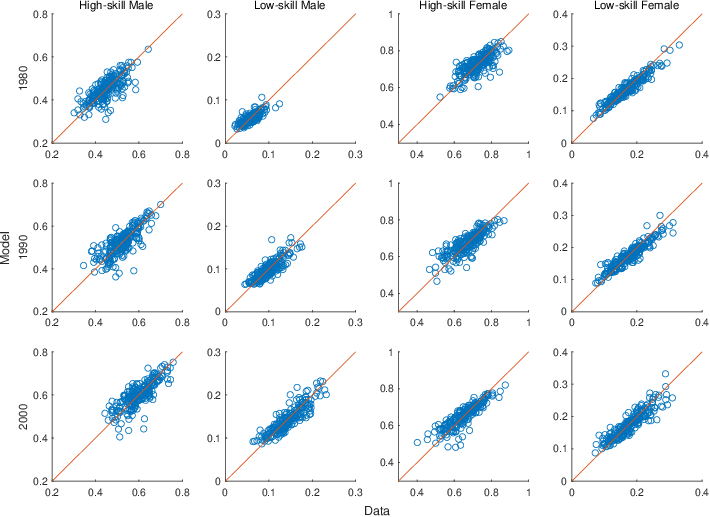} 
\caption{Model Fit---Scatter Plots of Data and Predicted PMH }
\label{fig:Model-Fit-PMH-2}

\vspace{1.0em} 
\raggedright
\scriptsize 
\noindent\parbox{1.0\textwidth}{
    \textit{Notes:} This figure compares the observed probability of marrying a high-skill spouse (PMH) with the model's prediction  by scatter plot, with a 45-degree reference line. Each data point represents the choice probabilities in a city.
    }
\end{minipage}
\end{figure}

Table \ref{fig:Model-Fit-L-1} and \ref{fig:Model-Fit-L-2} present the model fits related to individuals' location choices. Table \ref{fig:Model-Fit-PMH-2} shows scatter plots of the model's predicted PMH and the data.

% Figure

% Figure

\newpage
\section{Model with Birth Place Effects}

To better capture long-run migration costs and the empirical fact that people tend to live close to their hometown, we add birthplace into the benchmark model. Specifically, we let the indirect utility of a single individual $i$ living in city $m$ be given by 
\begin{align*}V_{imt}^{s} & =\ln W_{imt}-\zeta\ln R_{m,t}+f\left(\mathcal{A}_{mt};g_{i},e_{i}\right)+\varsigma_{im}+\sigma^{\epsilon}\epsilon_{it}^{e_{j}}+\sigma^{\nu}\nu_{imt},\end{align*}
where $\varsigma_{im}=\varsigma_{s\left(m\right)}\mathcal{I}\left(s\left(m\right)\ne s\left(i\right)\right)$, where $s\left(i\right)$ denotes the state of $i$'s hometown (birth place), $s\left(m\right)$ denotes the state of $i$'s current city, and $\varsigma_{s\left(m\right)}$ is a state-specific utility cost associated with moving out-of-town (out-of-state). Similarly, the indirect utility of individual $i$ married to partner $j$ is given by
\begin{align*}V_{imt}^{e_{j}} & =\ln\left(W_{imt}+W_{jmt}\right)-\zeta\ln R_{mt}-\ln\left(1+\chi\right)\\
 & \ \ \ \ +\mu_{t}^{e_{j}}\left(g_{i},e_{i}\right)+\tau_{mt}^{e_{j}}\left(g_{i},e_{i}\right)+f\left(\mathcal{A}_{mt};g_{i},e_{i}\right)+\varsigma_{im}+\sigma^{\epsilon}\epsilon_{it}^{e_{j}}+\sigma^{\nu}\nu_{imt}
\end{align*}

The addition of the birth place state effect does not change our estimation of local labor market, housing market, and marriage market parameters and only affects location choice estimation. To estimate location choice, we first compute 
\[
\widehat{u}_{imt}=\log\frac{\Pr\left(d_{imt}=1\right)}{\Pr\left(d_{im_{0}t}=1\right)}
\]
for each individual type that now includes gender, education, and birth place state. We then run the following regression
\[
\widehat{u}_{imt}=\widetilde{u}_{mt}^{g}+\varepsilon_{imt},
\]
where $\widetilde{u}_{mt}^{g}$ is a type-$g$ (gender, education) fixed effect. Doing so ``netted out'' the birth place state effect on each individual's utility for each location $m$. Finally, as in the estimation of the benchmark model, we estimate a time-differenced version of the following regression equation:
\begin{align*}
\widetilde{u}_{mt}^{g} & =\beta_{1}\mathbb{W}_{mt}^{g}+\beta_{2}^g\mathbb{A}_{mt}^{o}+a_{mt}^g,
\end{align*}
where we have further allowed the amenity preference parameter $\beta_{2}$ to depend on individual type. 

Table \ref{tab:result_bpl} gives the estimation result. 

% Table
\begin{table}[H]
\centering
\footnotesize
\caption{Location Choice Parameters}
\label{tab:result_bpl}
\begin{minipage}{0.99\textwidth}
\centering
\tabcolsep=6pt
\begin{tabular}{p{0.15\textwidth}*{5}{>{\centering}p{0.15\textwidth}}}
\toprule
\toprule
& $\beta_{1}$ & \multicolumn{4}{c}{$\beta_{2}$} \tabularnewline
 \cmidrule(lr){3-6}
& & mH & mL & fH & fL \tabularnewline
\midrule
Estimate & 7.636 & 0.070 & 0.127 & 0.104 & 0.053 \tabularnewline
S.E. & (3.970) & (0.056) & (0.060) & (0.065) & (0.048) \tabularnewline
\bottomrule
\bottomrule
\end{tabular}

\vspace{1.0em}
\raggedright
\scriptsize
\noindent\parbox{1.0\textwidth}{
\textit{Notes:} Standard errors are shown in parentheses. mH = male high-skill, mL = male low-skill, fH = female high-skill, and fL = female low-skill.
}
\end{minipage}
\end{table}
% end

\newpage
\section{Alternative Assortative Measures}
\label{ap:am}

We also explore the impacts of assuming a national marriage market on the estimated trends of assortative mating based on reduced-form measures that are commonly used in previous studies.

The likelihood ratio measure presented as below has been used by \citet{greenwood2014marry} and \citet{eika2019educational}. The intuition behind is to capture how observed matching of couples with the same education departs from random matching. Note that the denominator of the random matching probability controls for any changes in the marginal education distributions of both genders.\footnote{For multiple education categories, \citet{greenwood2014marry} calculate the likelihood ratio as the summation of probabilities of same-education matching over the summation of probabilities under random matching of each education category. \citet{eika2019educational} calculate weighted average of the likelihood ratio of each education category. Our measure follows the one calculated by \citet{greenwood2014marry}.} 
\begin{align*}
\text{LR} = \frac{\sum_{e_i,e_j} p(E^M=e_i, E^F=e_j)}{\sum_{e_i,e_j} p(E^M=e_i) \cdot p(E^F=e_j)    } 
\end{align*}

A more simple and statistical measure is the Pearson correlation coefficient, which measures the simple linear correlation between husband and wife's education level.\footnote{\citet{gihleb2020educational} points out the advantage of Pearson correlation with an nice example of perfect assortative mating, where the education levels of wives are just lower than ones of the husbands. In this example,  the likelihood ratio measure suggests no assortative mating as there is no match with same education levels. Pearson correlation measure takes this case into account.} While alternative correlation measures, such as the Kendall Tau or Spearman correlation coefficients, have been proposed to measure the correlation in rank, Pearson has been widely used as a benchmark for comparison and the results turn to to be similar among these correlation measures. One reminder is that the Pearson correlation coefficient also coincides with the measure of assortative mating based on the linear regression used by \citet{greenwood2014marry}, after properly controlling the variance of education of husband and wife.
\begin{align*}
    \rho = \frac{ \text{Cov}(E^M, E^F) } {  \sqrt{ {  \text{Var}(E^M)\text{Var}(E^F)  }   } }
\end{align*}

\begin{figure}[H] 
\centering 
\begin{minipage}[c]{0.99\textwidth} 
\centering
\includegraphics[height=5cm]{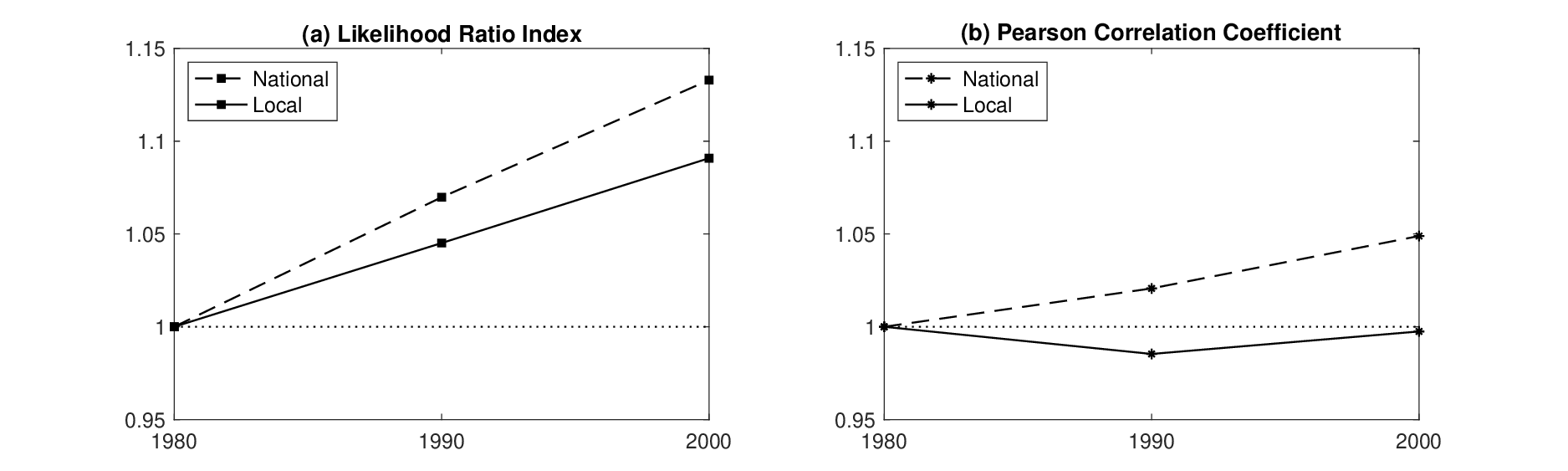}
\caption{Trends of Marital Assortativeness}
\label{fig:lrrho}

\vspace{1.0em} 
\raggedright
\scriptsize 
\noindent\parbox{1.0\textwidth}{
    \textit{Notes:} Left: Trends of assortative mating based on the likelihood ratio 
index. Right: Trends of assortative mating based on Pearson correlation coefficient.}
\end{minipage}
\end{figure}

Based on these two common measures of assortative mating, we estimate the national trends of assortative mating, with and without the national marriage assumption. As revealed by Figure \ref{fig:lrrho}, both the likelihood ratio index and Pearson correlation coefficient reveals a rising trend in assortative mating when local marriage markets are not considered. By allowing for different marital pools across cities, the trend measured by the likelihood ratio index drops significantly by 31.7\%. More surprisingly, Pearson correlation coefficient even reveals a picture that differs qualitatively: the trend of assortative mating has barely moved from 1980 to 2000.

% Figure

\newpage
\section{College Welfare Gap}

This section provides the results on the evolution of the college welfare gap by gender. Our results for males (Table \ref{table:wg_m}) are quantitatively comparable to the findings of \citet{moretti2013real} and \citet{diamond2016determinants}, which are based on samples of full-time employed male workers. Considering nominal and real wages respectively, \citet{moretti2013real} reveals 18 and 15 percentage points increase in the college welfare gap from 1980-2000, while \citet{diamond2016determinants} finds a widening gap of 21.8 and 19.4 percentage points. Similarly, we uncover a widening gap by 22.6 log points due to wage only and by 20.3 points due to real wages.

Comparing Tables \ref{table:wg_m} and \ref{table:wg_f}, we find that the welfare implications differ by several factors between males and females. First, the marriage market equilibrium significantly widened the college welfare gap among males but modestly dampened that among females. Second, nonpecuniary marital benefits significantly widened the college welfare gap among females but have more modest effects on males. Third, shifting education compositions significantly reduced the college welfare gap among females but widened the gap among males.

% Table
\begin{table}[H]
  \footnotesize
  \centering
  \begin{minipage}{0.99\textwidth}
   \caption{The College Welfare Gap from 1980 to 2000: Male}
   \tabcolsep=0.15cm
   \begin{tabular}{p{0.28\textwidth}>{\centering}p{0.1\textwidth}>{\centering}p{0.1\textwidth}>{\centering}p{0.1\textwidth}>{\centering}p{0.1\textwidth}>{\centering}p{0.1\textwidth}>{\centering\arraybackslash}p{0.1\textwidth}}
    \hline\hline
    & (1) & (2) & (3) & (4) & (5) & (6) \\
    \hline
    1980       & -0.053 & -0.053 & -0.053 & -0.053 &  -0.053    &  -0.053       \\
    2000       & 0.119  & 0.103  & 0.193  & 0.208  &   0.211    &   0.243    \\
    1980-2000  & 0.172  & 0.156  & 0.247  & 0.262  &   0.265    &   0.297    \\
    \hline
    Wage       & Yes    & Yes    & Yes    &   Yes     &   Yes    &   Yes    \\
    Rent       &     &     Yes   & Yes    &   Yes     &   Yes    &   Yes    \\

    Nonpecuniary Utility  &        &        &        &    Yes    &    Yes   &   Yes    \\
     Location Choice &        &        &        &        &  Yes    &   Yes    \\
           Education Composition &        &        &        &        &      &   Yes    \\         
    \hline
        Marriage Market Equilibrium       &   No     &    No    &   Yes     &   Yes     &   Yes    &   Yes    \\
      \hline  
   \end{tabular}
   
     \vspace{0.5em} 
   \scriptsize{ \textit{Notes:} This table presents changes in the college welfare gap for males from 1980 to 2000, based on counterfactual experiments sequentially allowing more factors to change over time. Column (1) allows only wages to change, while all other factors, including marital transfers, remain fixed at their 1980 levels. Column (2) adds changing rents to wages, without re-solving marriage market equilibria. Column (3) re-solves marriage market equilibria, allowing wages and rents to affect marital transfers. Column (4) further includes changes in nonpecuniary utilities. Column (5) incorporates changing location choice probabilities, and column (6) allows the marginal distributions of education levels to change as well. From column (3) onward, marital transfers adjust as part of the equilibria, rather than remaining fixed.}

   \label{table:wg_m}
  \end{minipage}
 \end{table}
% end

% Table
\begin{table}
  \footnotesize
  \centering
  \begin{minipage}{0.99\textwidth}
   \caption{The College Welfare Gap from 1980 to 2000: Female}
   \tabcolsep=0.15cm
   \begin{tabular}{p{0.28\textwidth}>{\centering}p{0.1\textwidth}>{\centering}p{0.1\textwidth}>{\centering}p{0.1\textwidth}>{\centering}p{0.1\textwidth}>{\centering}p{0.1\textwidth}>{\centering\arraybackslash}p{0.1\textwidth}}
    \hline\hline
    & (1) & (2) & (3) & (4) & (5) & (6) \\
    \hline
    1980       & -0.045 & -0.045 & -0.045 & -0.045 &  -0.045    &  -0.045       \\
    2000       & 0.116  & 0.103  & 0.085  & 0.332  &   0.334    &   0.220    \\
    1980-2000  & 0.161  & 0.148  & 0.131  & 0.377  &   0.379    &   0.265    \\
    \hline
    Wage       & Yes    & Yes    & Yes    &   Yes     &   Yes    &   Yes    \\
    Rent       &     &     Yes   & Yes    &   Yes     &   Yes    &   Yes    \\
    Nonpecuniary Utility  &        &        &        &    Yes    &    Yes   &   Yes    \\
     Location Choice &        &        &        &        &  Yes    &   Yes    \\
           Education Composition &        &        &        &        &      &   Yes    \\         
    \hline
        Marriage Market Equilibrium       &   No     &    No    &   Yes     &   Yes     &   Yes    &   Yes    \\
      \hline  
   \end{tabular}
   
    \vspace{0.5em} 
   \scriptsize{ \textit{Notes:} This table presents changes in the college welfare gap for females from 1980 to 2000, based on counterfactual experiments sequentially allowing more factors to change over time. Column (1) allows only wages to change, while all other factors, including marital transfers, remain fixed at their 1980 levels. Column (2) adds changing rents to wages, without re-solving marriage market equilibria. Column (3) re-solves marriage market equilibria, allowing wages and rents to affect marital transfers. Column (4) further includes changes in nonpecuniary utilities. Column (5) incorporates changing location choice probabilities, and column (6) allows the marginal distributions of education levels to change as well. From column (3) onward, marital transfers adjust as part of the equilibria, rather than remaining fixed.}
   
   \label{table:wg_f}
  \end{minipage}
 \end{table}
% end

\end{document}